\documentclass[fleqn, nofootinbib, twocolumn, notitlepage]{revtex4-1}

\usepackage{amsmath}    
\usepackage{mathtools}
\usepackage{bm}         
\usepackage{graphicx}   
\usepackage{grffile}
\usepackage{fancyvrb}   
\usepackage{color}      
\usepackage{subfigure}  
\usepackage{hyperref}   
\allowdisplaybreaks[1]
\usepackage[margin=0.7in]{geometry}
\def\d{\mathrm{d}}
\def\pd{\partial}
\def\b#1{\mathbf{#1}}

\usepackage{wrapfig}
\usepackage{titlesec}
\usepackage{amssymb}
\usepackage{upgreek}
\begin{document}
\title{Electrostatic $T$-matrix for a torus on bases of toroidal and spherical harmonics}
\author{Matt Majic}
\email{mattmajic@gmail.com}
\affiliation{The MacDiarmid Institute for Advanced Materials and Nanotechnology,
School of Chemical and Physical Sciences, \\
Victoria University of Wellington,
PO Box 600, Wellington 6140, New Zealand}

\begin{abstract}
Semi-analytic expressions for the static limit of the $T$-matrix for electromagnetic scattering are derived for a circular torus, expressed in bases of both toroidal and spherical harmonics. The scattering problem for an arbitrary static excitation is solved using toroidal harmonics and the extended boundary condition method to obtain analytic expressions for auxiliary $Q$ and $P$-matrices, from which the $T$-matrix is given by their division. By applying the basis transformations between toroidal and spherical harmonics, the quasi-static limit of the $T$-matrix block for electric multipole coupling is obtained. For the toroidal geometry there are two similar $T$-matrices on a spherical basis, for computing the scattered field both near the origin and in the far field. Static limits of the optical cross-sections are computed, and analytic expressions for the limit of a thin ring are derived.
\end{abstract}

\maketitle
\section{Introduction}

The $T$-matrix method is a semi-analytical tool for calculating the properties of electromagnetic or acoustic scattering by macroscopic particles \cite{waterman1965matrix,Mishchenko2002}. The incident and scattered electromagnetic fields are expanded on bases of orthogonal functions, e.g. spherical wavefunctions, and the $T$-matrix essentially outputs the series coefficients of the scattered field in terms of the known coefficients for the incident field. The $T$-matrix may be calculated via the extended boundary condition method (EBCM) which involves numerically evaluating surface integrals, however this approach is unstable for particles with highly non-spherical shape, and much work has been done on determining conditions for the $T$-matrix to be applicable \cite{farafonov2016analysis,kyurkchan1996singularities}. For the EBCM to work, the Rayleigh hypothesis is historically suggested as criteria, which is the assumption that the series of basis functions converge on the entire surface of the particle, although it is known that this is not necessary and weaker criteria have been suggested \cite{farafonov2016analysis}. The EBCM fails for certain convex particles, let alone multiply connected particles such as a torus.

Using other methods, scattering by a torus has been considered analytically in the long wavelength limit using toroidal harmonics, the partially separable solutions to Laplace's equation in toroidal coordinates. Toroidal harmonics are a relatively new tool in computational physics due to their complexity. Also, being only partially separable solutions makes toroidal harmonics difficult to apply to problems even involving the torus. For conducting tori, fairly simple series solutions can be obtained in the static limit, but for dielectric tori, the series coefficients are not explicit. The coefficients are calculated using a three step recurrence relation, with initial values given by a continued fraction.  
Here we use an alternative approach to this recurrence scheme by computing the $T$-matrix on a basis of toroidal harmonics. The series coefficients are then given by a matrix inversion. 
Some specific simulations include a dielectric torus in a uniform field \cite{love1972dielectric}, a conducting torus illuminated by a low frequency plane wave \cite{venkov2007low}, plasmon resonances \cite{dutta2008plasmonic} and electromagnetic trapping \cite{salhi2015toroidal}. A detailed computational analysis of quasistatic scattering by solid and layered tori is conducted in \cite{garapati2017poloidal}. Semi analytical results for low frequency dipole excitation of simple geometries - a conducting torus \cite{vafeas2016torusdipole},  spheroids \cite{vafeas2009spheroid} and two spheres \cite{vafeas2012twospheres} are obtained for the lowest three orders in frequency of the electric and magnetic fields. The problems were formulated using the natural harmonic functions for the geometry, but it is also practical and insightful to recast these solutions in terms of spherical harmonics. 

Spherical harmonics are far easier to compute, better known and more applicable than toroidal harmonics, so it is of interest to re-express this $T$-matrix on a basis of spherical harmonics - in particular, this allows computation of the long wavelength limit of the $T$-matrix for electromagnetic or acoustic scattering, which is usually expressed on a basis of spherical wavefunctions. For the spheroid, the low frequency scattering problem has been re-expressed with spherical harmonics and the electromagnetic $T$-matrix has been obtained to third order in size parameter \cite{TmatESA2017,majic2019quasistatic,majic2019approximate} by applying the series relationships between spherical and spheroidal harmonics. But the series relationships between spherical and toroidal harmonics are more complex and virtually unknown in the literature, exept for the low degrees: toroidal harmonics of degree zero, corresponding to the potential of rings of sinusoidal charge distributions, are known as series of spherical harmonics, and spherical harmonics corresponding to point charges and dipoles are known as series of toroidal harmonics. The relationships for all degrees and orders have been derived in a Russian paper from 1983 \cite{erofeenko1983addition}, although this does not appear to be well known except for one other paper \cite{shushkevich1998electrostatic}, which analysed the electrostatic interaction between a conducting torus and a partial spherical shell - they used the relationships to construct what is effectively the $T$-matrix for the conducting torus (eq. (32)), expressed on a spherical basis, as part of the kernel of an integral equation.\\

\textbf{Contents}
\begin{itemize}
	\item Section \ref{sec toroidal}:\\ 
	toroidal coordinates and harmonics are defined along with the charge distributions which generate these harmonics. \\
	\item Section \ref{sec toroidalTmat}: \\
	the scattering problem for a torus is formulated where the incident field is assumed to be expanded as a series of toroidal harmonics, and using the electrostatic null field equations, a toroidal $T$-matrix is derived which gives the scattered field also on a basis of toroidal harmonics. 
\item Section \ref{sec rels}:\\
 Relationships between spherical harmonics are presented, with a new simple recurrence for their expansion coefficients.
\item Section \ref{sec spherical}:\\ these relationships are used to re-expresses the $T$-matrix on a spherical harmonic basis.\\
\item  Section \ref{sec derived}: plots of the potential and electric field, and presents physical quantities including capacitance, polarizability, plasmon resonances and optical cross-sections. 
\item Section \ref{sec thin}: \\ 
Analytic asymptotic expressions for the $T$-matrix elements in the thin ring limit. 
\item Appendix: \\ 
Derivations of the expansions between spherical and toroidal harmonics along with some interesting recurrence relations, explicit forms, a generating function and symmetries of the expansion coefficients, aswell as symmetries between spherical and toroidal harmonics.
\end{itemize}

\section{Toroidal coordinates and harmonics} \label{sec toroidal}
Firstly we introduce spherical coordinates $(r,\theta,\phi)$, and cylindrical coordinates $(\rho,z,\phi)$. The toroidal coordinate system also uses a parameter $a$ which defines the radius of the "focal ring" centred at the origin. Toroidal coordinates $(\xi,\eta,\phi)$ are then defined as 
\begin{align}
\eta&=\text{sign}(z)\arccos\frac{r^2-a^2}{\sqrt{(r^2+a^2)^2-4\rho^2a^2}},\\
\xi&=\frac{1}{2}\log\frac{(\rho+a)^2+z^2}{(\rho-a)^2+z^2}, \\
\beta&=\cosh\xi,
\end{align} 
with $\eta\in[-\pi,\pi],~\xi\in[0,\infty),~\beta\in[1,\infty)$. $\beta$ corresponds to the torus size ($\beta=1$ is a tight torus covering all space and $\beta=\infty$ is the focal ring) and $\eta$ relates to the angle around the minor axis.

Laplace's equation is partially separable in toroidal coordinates, so that the solutions have an additional prefactor. The toroidal harmonics are defined for integer $n,m$ as
\begin{align}
\psi_n^{mc} &= \Delta Q_{n-1/2}^m(\beta)\cos n\eta~ e^{im\phi}, \\
\psi_n^{ms} &= \Delta Q_{n-1/2}^m(\beta) \sin n\eta~ e^{im\phi}, \\
\Psi_n^{mc} &= \Delta P_{n-1/2}^m(\beta)\cos n\eta~ e^{im\phi}, \\
\Psi_n^{ms} &= \Delta P_{n-1/2}^m(\beta)  \sin n\eta~ e^{im\phi}, \\ 
\text{with }\Delta&=\sqrt{2(\beta-\cos\eta)},
\end{align}
where $P_{n-1/2}^m$ and $Q_{n-1/2}^m$ are the Legendre functions of the first and second kinds.
Using the superscript $v$ to denote $c$ or $s$, we will call $\Psi_n^{mv}$ the ``ring toroidal harmonics" and $\psi_n^{mv}$ the ``axial toroidal harmonics" since they are singular on the focal ring and $z$-axis respectively. There are two kinds of angular solutions - $v=c~(\cos n\eta)$ which are symmetric about the meridian plane and $v=s~(\sin n\eta)$ which are antisymmetric. Compare this to spherical harmonics which have angular solutions $P_n^m(\cos\theta)$ and $Q_n^m(\cos\theta)$ where the second type of angular solutions $Q_n^m(\cos\theta)$ are discarded due to their singularity at the poles of the sphere. Toroidal harmonics also are non-zero for $m>n$ - as shown in the next section, this occurs (with $n=0$ for example) for a single ring with charge distribution $\cos(m\phi)$. So all $\Psi_n^{mv}$ and $\psi_n^{mv}$ for integer $n,m$ and $v=c,s$ are physically applicable solutions - smooth on the torus surface. 
And toroidal harmonics of negative $n$ or $m$ may be defined simply through
\begin{align}
P_{-n-1/2}^m&=P_{n-1/2}^m \\
Q_{-n-1/2}^m&=Q_{n-1/2}^m \\
P_{n-1/2}^{-m}&=\frac{\Gamma(n-m+\frac{1}{2})}{\Gamma(n+m+\frac{1}{2})}P_{n-1/2}^m \\
Q_{n-1/2}^{-m}&=\frac{\Gamma(n-m+\frac{1}{2})}{\Gamma(n+m+\frac{1}{2})}Q_{n-1/2}^m. \label{PQ -m}
\end{align}

\subsection{Charge distributions of ring toroidal harmonics $\Psi_n^{mv}$} \label{sec ring charge}
In order to give insight, we investigate the charge distributions that create toroidal harmonics. $\Psi_n^{mv}$ are finite everywhere except $\beta=\infty$, on the focal ring. We want to express the functions as integrals of charge distributions on this focal ring, weighted by the inverse distance between a point on the focal ring $\b r'$ and an observation point $\b r$. We start with $n=0$, looking at $\Psi_0^{mc}=\Delta P_{-1/2}^m(\beta)e^{im\phi}$ (note that $\Psi_0^{ms}=0$). The Legendre functions approaching the ring behave as
\begin{align}
P_{-1/2}^m(\beta\rightarrow\infty)\rightarrow \frac{(2m-1)!!}{(-2)^m\pi}\sqrt{\frac{2}{\beta}}\log\beta
\end{align}
while $\Delta\rightarrow\sqrt{2\beta}$. Approaching the focal ring, $1/\beta$ becomes equal to the distance $d$ from the ring. Compare this to approaching infinitely close to a line source with arbitrary charge distribution, where the potential goes as $2\log d$ if the charge has unit density at the point of closest approach. The distance $|\b r'-\b r|$ in cylindrical coordinates is $|\b r'-\b r|=\sqrt{\rho^2+a^2-2\rho a\cos(\phi-\phi')+z^2}$. And the charge distribution must be proportional to $e^{im\phi}$, so we deduce that
\begin{align}
\Psi_0^{mc}
&=\frac{(2m-1)!!}{(-2)^m\pi}\int_0^{2\pi}\!\!\frac{e^{im\phi'}a\d\phi'}{\sqrt{r^2+a^2-2\rho a\cos(\phi\!-\!\phi')}}. \label{int Psi n=0}
\end{align}
We should also check the limit as $r\rightarrow\infty$. Here $\beta\rightarrow1$, $P_{n-1/2}\rightarrow1,~~ P_{n-1/2}^{m>0}\rightarrow0$, and $\Delta \rightarrow 2a/r$, so we can rule out the possibility of sources at $r=\infty$.\\

\begin{figure}
\includegraphics[scale=.52]{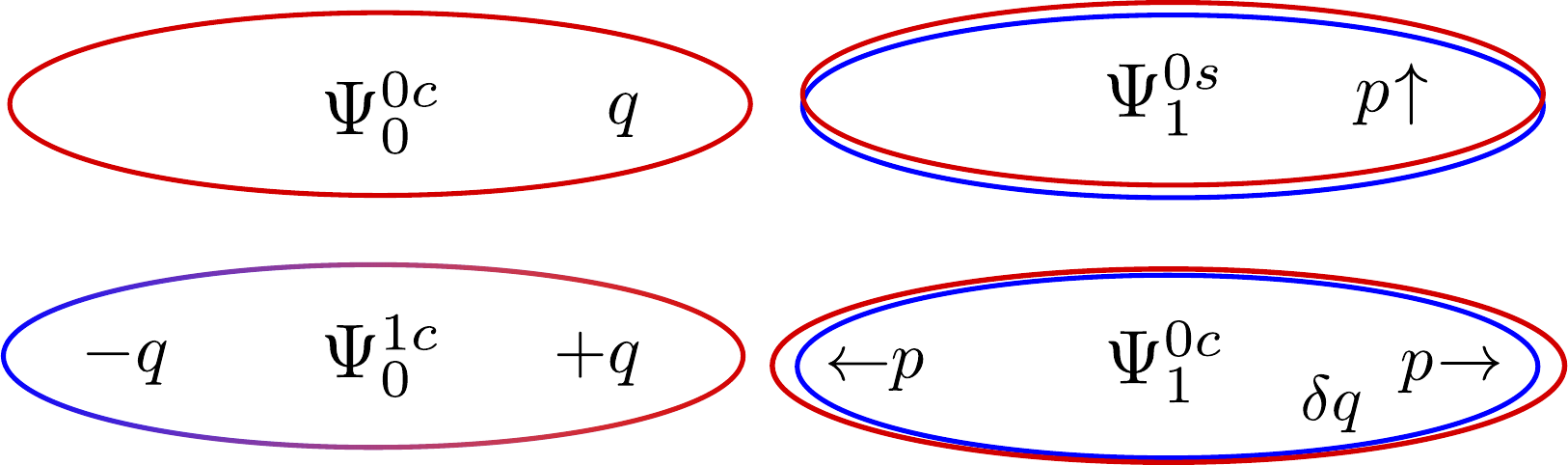}
\caption{Representation of the source charge configurations for the low order ring toroidal harmonics. Red and blue represent positive and negative charge $q$, and dipole moments are denoted $p$. For $\Psi_1^{0c}$ the slightly larger radius of the outer ring creates an infinitesimal charge $\delta q$. }\label{ringcharges}
\end{figure}

The harmonics for $n=1$ may be generated by application of the operators $\pd_z=\pd/\pd z$ and $r\pd_r$ (see appendix \ref{toroidal vs spherical appendix} for details), so we apply these to the integral expression \eqref{int Psi n=0}, first $r\pd_r$:
\begin{align}
\Psi_1^{mc}=&-\frac{2r\pd_r+1}{m-1/2} \Psi_0^{mc} \nonumber\\
=&\frac{(2m-3)!!}{(-2)^m\pi}\int_0^{2\pi}\!\!\frac{a (a^2-r^2)e^{im\phi'}\d\phi'}{(r^2+a^2-2\rho a\cos(\phi\!-\!\phi'))^{3/2}} \label{int Psi n=1 c}
\end{align}
The charge distribution is two oppositely charged rings on the $xy$-plane, one with an infinitesimally greater radius than the other. This produces an infinitesimal charge imbalance - a net monopole moment, which is seen in the spherical-toroidal expansions \eqref{cos_toroidal_spherical}.\\
And for $\Psi_1^{ms}$:
\begin{align}
\Psi_1^{ms}=&\frac{-a\pd_z}{m-1/2}\Psi_0^{mc}\nonumber\\
=&\frac{(2m-3)!!}{(-2)^m\pi}\int_0^{2\pi}\!\!\frac{-2a^2 ze^{im\phi'}\d\phi'}{(r^2+a^2-2\rho a\cos(\phi\!-\!\phi'))^{3/2}} \label{int Psi n=1 s}
\end{align}
This is the potential of two oppositely charged rings with an infinitesimal separation in the $z$-direction. These charge distributions are represented in figure \ref{ringcharges}.

For higher $n$ the harmonics can be generated by repeated application of the operator $r\pd_r$. We leave the details for appendix \ref{toroidal vs spherical appendix}, but state that the toroidal harmonics follow a recurrence relation \eqref{rdrQ} involving $r\pd_r$, and this recurrence is also satisfied by coefficients $c_{nk}^m$, $s_{nk}^m$ relating toroidal and spherical harmonics (see \eqref{c s rec}). We can then deduce the differential operators that generate the $n^{th}$ order harmonic from the $0^{th}$ or $1^{st}$ order harmonics:
\begin{align}
\Psi_n^{mc}&=\frac{(-)^n}{2}c_n^m(r\pd_r)\Psi_0^{mc} \label{rec toroidal cos}\\
\Psi_n^{ms}&=\frac{(-)^{n+1}}{4}\frac{s_n^m(r\pd_r)}{r\pd_r+1}\Psi_1^{ms} \label{rec toroidal sin}
\end{align}
where $c_n^m(r\pd_r)$ is equal to $c_{nk}^m$ with $k\rightarrow r\pd_r$ and similarly for $s_n^m$. \eqref{rec toroidal sin} starts from $n=1$ because $\Psi_0^{ms}=0$. Both $c_n^m(x)$ and $s_n^m(x)/(x+1)(x)$ are degree $n$ polynomials  in $x$.

The corresponding integral expressions for $\Psi_n^{mv}$ are given in operator form by applying \eqref{rec toroidal cos} and \eqref{rec toroidal sin} to the integral expression for $\Psi_0^{mc}$ \eqref{int Psi n=0} and $\Psi_1^{ms}$ \eqref{int Psi n=1 s}. However, explicit evaluations of these derivatives do not appear to reveal any simple patterns. The integral expressions for $\Psi_n^{mv}$ have been checked numerically up to $n=3$.

\subsection{Charge distributions of axial toroidal harmonics $\psi_n^{mv}$}
We will use heuristic arguments to derive the line source distributions for  $\psi_n^{mv}$, which are singular on the $z$ axis. First consider $m=0$.
As in the previous section, we use the fact that for an arbitrary line charge distribution, close enough to the line, the potential will only depend on the magnitude of the charge distribution at the point of approach. So we can determine the line charge distribution by matching it to the limit of the potential as it approaches the line. The exact proportionality is obtained by considering that the potential near a line source with unit magnitude goes as $2\log(\rho)$. The behaviour of the components of the toroidal harmonics as $\rho\rightarrow0$ are, using $v=z/a$:
\begin{align}
\lim_{\rho\rightarrow0} \Delta =& \frac{2}{\sqrt{v^2+1}}\\
\lim_{\rho\rightarrow0} \eta   =& \text{sign}(v)\text{acos}\frac{v^2-1}{v^2+1} \equiv \eta'\\
\lim_{\beta\rightarrow1} Q_{n-1/2}(\beta) =& \log\frac{1}{\sqrt{\beta-1}} \\
\lim_{\rho\rightarrow0} \frac{1}{\sqrt{\beta-1}}  =& \frac{a}{\rho}(v^2+1).
\end{align}
Then the line source distribution of $\psi_n^{c}+i\psi_n^{s}$ must be $a e^{in\eta'}/\sqrt{v^2+1}$ (using complex notation to deal with $\psi_n^c$ and $\psi_n^s$ simultaneously):
\begin{align}
\psi_n^{c}+i\psi_n^{s}=\int_{-\infty}^\infty  \frac{a~e^{in\eta'}\d v}{\sqrt{v^2+1}\sqrt{\rho^2+(z-av)^2}}. \label{line}
\end{align}
And for $\rho\rightarrow\infty$, we have $\psi_n^v\rightarrow0$ and there is no contribution from sources at $\rho\rightarrow\infty$. This is unlike the spherical harmonics of the second kind $r^nQ_n(\cos\theta)$, whose charge distributions are the difference between a line source that produces an infinite potential and a sum of multipoles at infinity of infinite strength (work currently in progress). 

Note that function $\text{sign}(v)$ makes the charge distribution continuous, and that it can also be expressed in terms of Chebyshev polynomials of the first and second kinds $T_n$ and $U_n$:
\begin{align}
e^{in\eta'}=T_n\bigg(\frac{v^2-1}{v^2+1}\bigg) + \frac{2i v}{v^2+1}U_{n-1}\bigg(\frac{v^2-1}{v^2+1}\bigg). 
\end{align}
For $m>0$, the charge distributions are multi-line, which can be deduced from the $e^{im\phi}$ dependence. For example $\psi_n^{1v}$ has a charge distribution of two line sources infinitely close together but of opposite charge. The behaviour of the Legendre functions $Q_{n-1/2}^m(\beta)$ is:
\begin{align}
\lim_{\rho\rightarrow0} Q_{n-1/2}^m(\beta) 
=& \frac{(-)^m}{2}(m-1)!\left(\frac{a(v^2+1)}{\rho}\right)^m. \label{limitQnm}
\end{align}
So for $m>0$ we have, $\psi_n^{mv}\propto\rho^{-m}$ as $\rho=0$. 

And in the limit $\rho\rightarrow\infty$, we have $\psi_n^{mv}\propto\rho^{m-1}$. Despite the divergence for $m>1$, there is no contribution from sources at $\rho=\infty$, which can be explained as follows. If sources at $\rho=\infty$ existed, then we can split $\psi_n^{mv}$ into its contributions from charges on the $z$-axis and at $\rho=\infty$. The potential due to charges at $\rho=\infty$, being finite at the origin, can then be expressed as a series of regular spherical harmonics $\hat S_n^m= r^nP_n^m(\cos\theta)e^{im\phi}$ of the same $m$ as $\psi_n^{mv}$. However, $\hat{S}_n^m(\rho\rightarrow\infty)\propto\rho^k$ where $k\geq m$, so it is impossible to express the $\rho^{m-1}$ dependence as a series of $\hat{S}_n^m$. Therefore there cannot exist charges at $\rho=\infty$. 

We can compare $\psi_n^{mv}$ to the potential near an $m$-fold line charge distribution which similarly goes as $\rho^{-m}$ as $\rho\rightarrow0$, with integral kernel $|\b r-\b r'|^{-2m-1}$. \eqref{limitQnm} for $m>0$ also shows a non-constant $z$-dependence $(v^2+1)^m$. Putting this together gives
\begin{align}
\psi_n^{mc}+i\psi_n^{ms}=&(2m-1)!! a\bigg(\frac{-a\rho}{2}\bigg)^m e^{im\phi} \nonumber\\&\times \int_{-\infty}^\infty  \frac{(v^2+1)^{m-1/2}e^{in\eta'}}{(\rho^2+(z-av)^2)^{m+1/2}}\d v  \label{line}
\end{align}
Andrews (\cite{Andrews2006}- eq. 27) proved \eqref{line} via direct integration for all $m$, and $n=0$ (in his notation $n\leftrightarrow m$).

\section{$T$-matrix for a torus on a toroidal harmonic basis} \label{sec toroidalTmat}
\begin{figure}
\includegraphics[scale=1,trim={0cm .3cm -1cm 0cm},clip]{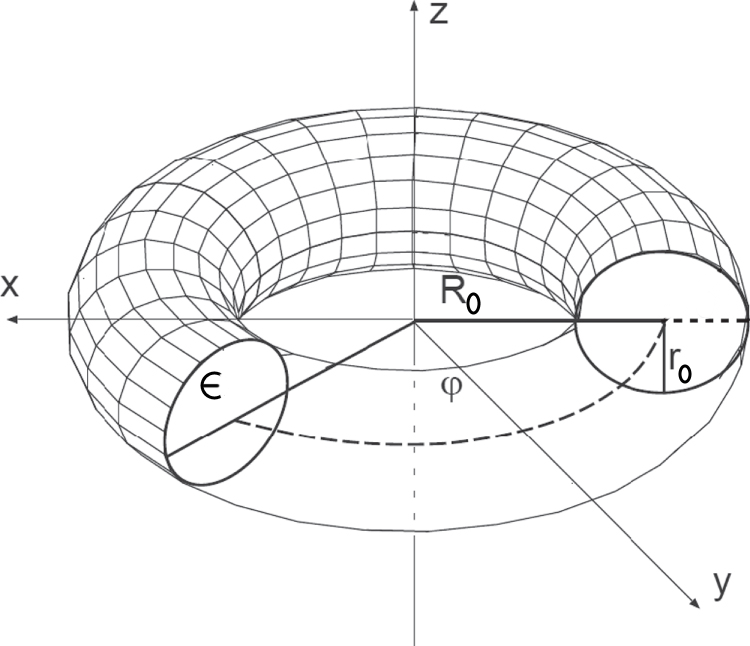}
\caption{Above: Parameters defining the torus including relative permittivity to the surroundings $\epsilon$. Below: relationships between toroidal geometry and the separable toroidal coordinate system.}
\includegraphics[scale=.325,trim={0cm .5cm 0cm 0cm},clip]{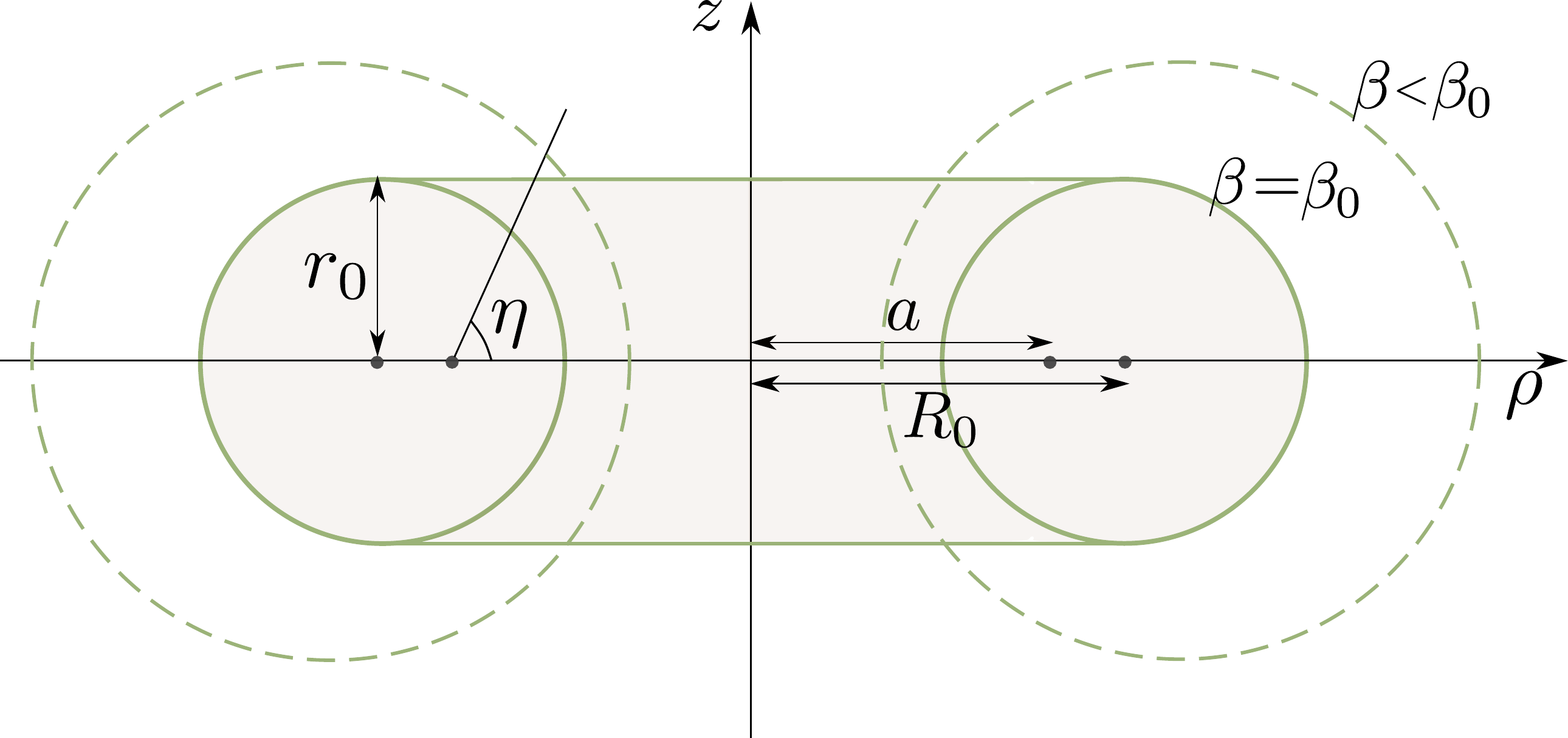}
\label{torus}
\end{figure}
Consider a circular torus with major radius $R_0$, minor radius $r_0$ and relative permittivity to the surroundings $\epsilon$ as shown in figure \ref{torus}. The surface is defined by the aspect ratio $\beta=\beta_0=R_0/r_0$, and the focal ring radius is $a=\sqrt{R_0^2-r_0^2}$. The torus is excited by an arbitrary external electric potential $V_e$, and we want to determine the scattered potential $V_s$ (where $V_o=V_e+V_s$), and internal potential $V_i$. 
The boundary conditions are:
\begin{align}
V_i=V_o, \qquad \epsilon\frac{\pd V_i}{\pd\beta}  = \frac{\pd V_o}{\pd\beta}, \qquad \text{at }\beta=\beta_0. \label{BCs}
\end{align}
The fields are assumed to be expanded in terms of toroidal harmonics as 
\begin{align}
V_e&=\sum_{n=0}^\infty \sum_{m=-\infty}^\infty\sum_{v=c,s} a_n^{mv}\psi_n^{mv}, \label{Vex tor}\\
V_i&=\sum_{n=0}^\infty \sum_{m=-\infty}^\infty\sum_{v=c,s} b_n^{mv}\psi_n^{mv}, \label{Vin tor}\\
V_s&=\sum_{n=0}^\infty \sum_{m=-\infty}^\infty\sum_{v=c,s} c_n^{mv}\Psi_n^{mv}. \label{Vsc tor}
\end{align}
In this section we derive the matrix relations between the incident, scattered, and internal expansion coefficients. 

In general the boundary conditions are difficult to solve, and may lead to an inhomogeneous three term recurrence relation with initial conditions given by an infinite continued fraction \cite{love1972dielectric}. Instead we will use the boundary integral equation formulation to find an analogue of the $T$-matrix in toroidal coordinates. The boundary integral equations are derived from Green's second identity on the torus surface $S$ \cite{farafonov2014rayleigh}:
\begin{align}
\frac{\epsilon-1}{4\pi}\int_S\!\frac{\pd V_i(\b r')}{\pd n'} \frac{1}{|\b r-\b r'|}\d S' = 
\begin{cases}
V_e(\b r)\!-\!V_i(\b r) & \b r\in V \\
-V_s(\b r) & \b r\notin V.  
\end{cases} \label{null}
\end{align}
where $\pd/\pd n'$ is the derivative with respect to the surface normal and $\d S$ is the infinitesimal surface element. In the $T$-matrix approach, Green's function $1/|\b r-\b r'|$ is expanded as a series of separable harmonics; here we use toroidal harmonics:
\begin{align}
\frac{1}{|\b r-\b r'|}=\frac{1}{2\pi a}&\sum_{m=-\infty}^\infty \sum_{n=0}^\infty\sum_{v=c,s}\varepsilon_n (-)^m \nonumber\\
&\times \begin{cases}
\psi_n^{mv}(\b r)\Psi_n^{-m,v}(\b r') \quad & \b r \in V\\ 
\Psi_n^{mv}(\b r)\psi_n^{-m,v}(\b r') & \b r \notin V
\end{cases} \label{GFexp} \\
\text{with }\varepsilon_n&=\begin{cases}1\quad   n=0 \\
2\quad   n>0 \end{cases}. \nonumber
\end{align}
We now substitute \eqref{GFexp} and the expansions of the potentials into the integral equation\eqref{null}, and equate the coefficients of the toroidal harmonics. For example, equating the coefficients of $\psi_n^{mc}(\b r)$ for $\b r$ inside the torus leads to (dropping the integration primes):
\begin{align}
\varepsilon_n&(-)^m\frac{\epsilon-1}{8\pi^2a}\sum_{k=0}^\infty\int_S \frac{\pd}{\pd n}\{\Delta Q_{k-1/2}^m(\beta_0)[b_k^{mc}\cos(k\eta)+\nonumber\\&b_k^{ms}\sin(k\eta)]\}\Delta P_{n-1/2}^{-m}(\beta_0)\cos(n\eta)\d S = a_n^{mc}-b_n^{mc}. \label{int1}
\end{align}
The sum over $m$ has been omitted because the torus is rotationally symmetric and only terms with the same $m$ in the expansions of the $V_i$ and $G$ survive the integration. This decouples the problem for each $m$. Furthermore, the integration is over an even interval of $\eta$ so $\sin(k\eta)\cos(n\eta)$ integrates to zero, decoupling the sine and cosine expansion coefficients.
The relationships between the expansion coefficients can be formulated with infinite dimensional matrices. For each $m\in\mathbb{Z}, ~ v=c,s$, the problem is solved by the following matrix relations: 
\begin{align}
&\bar{\b a}\!=\!\bar{\b Q}\bar{\b b}, \quad \bar{\b c}\!=\!\bar{\b P}\bar{\b b}, \quad \bar{\b c}\!=\!\bar{\b T}\bar{\b a} \label{Ttoroidal}
\end{align}
which are analogous to the P, Q, $T$-matrices in the $T$-matrix method for scattering of waves. $\bar{\b a}$ are $1\times N$ ($N$ being the numerical truncation order) column vectors containing the elements $a_n^{mv}$ for fixed $m,v$. When convenient the superscripts $m$ and $v$ will be omitted.\\
For a conducting torus, $\epsilon\rightarrow\infty$ and the $T$-matrix $\bar{\b T}=\bar{\b T}^\infty$ is diagonal:
\begin{align}
\bar{T}^\infty_{nk}=-\frac{Q_{n-1/2}^m(\beta_0)}{P_{n-1/2}^m(\beta_0)}\delta_{nk} \label{conducting solution}
\end{align}
which is equivalent to well known solutions for conducting tori, see for example \cite{Scharstein2005}.\\
For general $\epsilon$, the integral equations \ref{null} allow us to calculate $\bar{\b P}$ and $\bar{\b Q}$, while $\bar{\b T}$ is obtained from
\begin{align}
\bar{\b T}=\bar{\b P}\bar{\b Q}^{-1}. \label{TPQ}
\end{align}
We now look for analytic expressions for the matrix elements. In toroidal coordinates the normal derivative and surface element are
\begin{align}
\frac{\pd}{\pd n}=\frac{-\Delta^2\sinh\xi}{2a}\frac{\pd}{\pd\beta}, \qquad 
\d S= \frac{4a^2\sinh\xi}{\Delta^4}\d\eta\d\phi.
\end{align}
Putting these into \eqref{int1} and simplifying leads to an integral expression for $\bar{\b Q}^{c}$: 
\begin{align}
\bar{Q}_{nk}^{mc}=&\delta_{nk}-\frac{\epsilon-1}{2\pi}P_{n-1/2}^{-m}(\beta_0)\varepsilon_n \nonumber\\
&\times\bigg[ \frac{\pd Q_{k-1/2}^m(\beta_0)}{\pd\beta_0}\int_{-\pi}^\pi \cos(n\eta)\cos(k\eta)\d\eta \nonumber\\
& +\frac{1}{2}
Q_{k-1/2}^m(\beta_0)\int_{-\pi}^\pi \frac{\cos(n\eta)\cos(k\eta)}{\beta_0-\cos\eta}\d\eta \bigg].  \label{int2}
\end{align}
The first integral is simple:
\begin{align}
\int_{-\pi}^\pi \cos(n\eta)\cos(k\eta)\d\eta =2\pi\frac{\delta_{nk}}{\varepsilon_n}, \label{cosine int}
\end{align}
and the second integral evaluates to 
\begin{align}
\int_{-\pi}^\pi \frac{\cos(n\eta)\cos(k\eta)}{\cosh\xi_0-\cos\eta}\d\eta = \pi \frac{e^{-|n-k|\xi_0}+e^{-(n+k)\xi_0}}{\sinh\xi_0}.
\label{Tmat int cos}
\end{align}
See appendix \ref{int proof} for proof. For $\bar{\b Q}^{s}$ the integrals are
\begin{align}
\int_{-\pi}^\pi \sin(n\eta)\sin(k\eta)\d\eta =\pi\delta_{nk}(1-\delta_{n0}), \label{sine int}
\end{align}
and 
\begin{align}
\int_{-\pi}^\pi \frac{\sin(n\eta)\sin(k\eta)}{\cosh\xi_0-\cos\eta}\d\eta = -\pi  \frac{e^{-|n-k|\xi_0}-e^{-(n+k)\xi_0}}{\sinh\xi_0} . \label{Tmat int sin}
\end{align}
The matrix elements are then

\begin{align}
&\bar{Q}_{nk}^{mc}=\delta_{nk}\!- (\epsilon\!-\!1)\sinh^2\!\xi_0 P_{
\! n-1/2}^{-m}(\beta_0)\bigg[\pd_{\beta_0}Q_{n-1/2}^m(\beta_0)\delta_{nk}\nonumber\\
&+ \frac{\varepsilon_n}{2}Q_{k-1/2}^m(\beta_0)\frac{e^{-|n-k|\xi_0}+e^{-(n+k)\xi_0}}{2\sinh\xi_0} \bigg], \label{Qc}\\
&\bar{Q}_{nk}^{ms}=\delta_{nk}\!- (\epsilon\!-\!1)\sinh^2\!\xi_0 P_{\! n-1/2}^{-m}(\beta_0)\bigg[\pd_{\beta_0}Q_{n-1/2}^m(\beta_0)\delta_{nk}\nonumber\\
&\times(1-\delta_{n0}) - Q_{k-1/2}^m(\beta_0)\frac{e^{-|n-k|\xi_0}-e^{-(n+k)\xi_0}}{2\sinh\xi_0} \bigg]. \label{Qs}
\end{align}

A similar derivation shows that $\bar{\b P}$ is related to $\bar{\b Q}$ by
\begin{align}
\bar{\b P}&=\bar{\b T}^\infty(\bar{\b Q}-\b I)\label{P algebraic} 
\end{align}
for all $m,v$. Hence the $T$-matrix can also be expressed as
\begin{align}
\bar{\b T}=\bar{\b T}^\infty(\b I-\bar{\b Q}^{-1}). \label{ThatP}
\end{align}
We suspect this form applies to the $T$-matrix for any particle who's geometry is a coordinate of a coordinate system with partially-separable solutions to Laplace's equation, including bispherical coordinates (in preparation). $\bar{\b T}$ may be calculated via \eqref{TPQ} or \eqref{ThatP}; both give the same numerical accuracy. To avoid possible numerical instability in inverting $\bar{\b Q}^{m=0,c}$, the matrix should be transposed, inverted then transposed back.\\

The static limit $T$-matrix for any scatterer is symmetric when expressed on a basis of normalised spherical harmonics or wave functions. This is proved in \cite{farafonov2015EBCM} by equating solutions from two surface integral equations similar to \eqref{null}, given in \cite{farafonov2014rayleigh}. The proof used a basis of normalised spherical harmonics but could be easily done with any orthonormal basis. The basis functions $\Psi_{n}^{mv}, \psi_{n}^{mv}$ are orthogonal but not normalised - they have a norm of $\Gamma(n+m+\frac{1}{2})/\Gamma(n-m+\frac{1}{2})/\varepsilon_n$. Hence the toroidal $T$-matrix has the following symmetry property:
\begin{align}
\bar T_{kn}^{mv}=\frac{\varepsilon_k\Gamma(k-m+\frac{1}{2})\Gamma(n+m+\frac{1}{2})}{\varepsilon_n\Gamma(k+m+\frac{1}{2})\Gamma(n-m+\frac{1}{2})}\bar T_{nk}^{mv}. \label{symmetry}
\end{align}


\subsection{Comparison to recurrence approach}
Another way to solve electrostatic problems for the dielectric torus is to apply the boundary conditions in differential form \eqref{BCs} directly to the series expansions for the potentials (\ref{Vex tor}-\ref{Vsc tor}), to obtain a recurrence relation for the coefficients. This has been done for a uniform electric field \cite{love1972dielectric}, and point charge \cite{kuyucak1998analytical}. For arbitrary excitation, we get
\begin{align}
&c_{n+1}^{mv}\Lambda_{n+1}^m - c_n^{mv}\big(2\beta_0 \Lambda_n^m+(\epsilon-1)P_{n-1/2}^m \big) + c_{n-1}^{mv}\Lambda_{n-1}^m \nonumber\\
&= (\epsilon-1)\big[- a_{n+1}^{mv}Q_{n+1/2}^{m\prime} + a_n^{mv}(Q_{n-1/2}^m+2\beta_0 Q_{n-1/2}^{m\prime}) \nonumber\\
&~~~~- a_{n-1}^{mv}Q_{n-3/2}^{m\prime} \big] \nonumber\\
&\text{where }~~ \Lambda_n^m=\epsilon Q_{n-1/2}^{m\prime}(\beta_0)\frac{P_{n-1/2}^m(\beta_0)   }{Q_{n-1/2}^m(\beta_0)}  - P_{n-1/2}^{m\prime}(\beta_0). \label{rec approach}
\end{align}
The difficulty in computing the coefficients using this scheme is finding the initial values; these are expressed as a series of products of continued fractions.
Nevertheless, we can compare this to the $T$-matrix approach. Combining recurrence \eqref{rec approach} with the definition of the $T$-matrix, \eqref{Ttoroidal}, we find
\begin{align}
&\sum_{k=0}^\infty \bigg[ \bar T_{n+1k}^{mv}\Lambda_{n+1}^m - \bar T_{nk}^{mv}\big(2\beta_0 \Lambda_n^m+(\epsilon-1)P_{n-1/2}^m \big) \nonumber\\
&\quad+ \bar T_{n-1k}^{mv}\Lambda_{n-1}^m \bigg]a_k^{mv} \nonumber\\
&= (\epsilon-1)\big[ - a_{n+1}^{mv}Q_{n+1/2}^{m\prime} + a_n^{mv}(Q_{n-1/2}^m+2\beta_0 Q_{n-1/2}^{m\prime})\nonumber\\
&~~~~- a_{n-1}^{mv}Q_{n-3/2}^{m\prime} \big] 
\end{align}
which must hold for any set of coefficients $a_n^{mv}$. In particular if we choose an excitation with $a_n^{mv}=\delta_{np}$ for some $p$, a recurrence for the elements of the toroidal $T$-matrix can be found:
\begin{align}
& \bar T_{n+1p}^{mv}\Lambda_{n+1}^m - \bar T_{np}^{mv}\big(2\beta_0 \Lambda_n^m+(\epsilon-1)P_{n-1/2}^m \big) + \bar T_{n-1p}^{mv}\Lambda_{n-1}^m \nonumber\\
&= (\epsilon-1)\big[\delta_{np}(Q_{p-1/2}^m+2\beta_0 Q_{p-1/2}^{m\prime})  \nonumber\\
&~~~~- (\delta_{n+1p}+\delta_{n-1p}) Q_{p-1/2}^{m\prime} \big].
\end{align}
Which has been used as a check for the $T$-matrix.

\section{Relationships between spherical and toroidal harmonics} \label{sec rels}
In order to express the $T$-matrix on a basis of spherical harmonics, series relationships between spherical and toroidal harmonics are needed. 
We define the regular and irregular solid spherical harmonics as:
\begin{align}
\hat S_n^m&= \left(\frac{r}{a}\right)^n P_n^m(\cos\theta)e^{im\phi}, \\ 
S_n^m&= \left(\frac{a}{r}\right)^{n+1}P_n^m(\cos\theta)e^{im\phi}. \label{sph funcs}
\end{align}

In the appendix we derive the following linear relationships between spherical and toroidal harmonics:
\begin{align}
\Psi_n^{mc} &= \sum_{k=m}^\infty c_{nk}^mP_k^{-m}(0)
\begin{dcases}
(-)^n\hat S_k^m \quad &r<a\\
S_k^m \quad &r>a\\
\end{dcases} \label{cos_toroidal_spherical}\\
\Psi_n^{ms} &= \sum_{k=m}^\infty s_{nk}^mP_{k+1}^{-m}(0)
\begin{dcases}
(-)^n\hat S_k^m \quad &r<a\\
-S_k^m \quad &r>a
\end{dcases} \label{sin_toroidal_spherical}\\
&\hat S_n^m=\frac{1}{\pi}
\begin{dcases}\!P_n^m(0)
\!\sum_{k=0}^\infty \frac{\varepsilon_k}{2} c_{kn}^{-m}\psi_k^{mc} 
~~ &n\!+\!m\text{ even} \\
\!-P_{n+1}^m(0)\!\sum_{k=1}^\infty\! s_{kn}^{-m}\psi_k^{ms}  &n\!+\!m\text{ odd}
\end{dcases} \label{reg_spherical_toroidal} \\
&S_n^m=\frac{1}{\pi}
\begin{dcases}
\!P_n^m(0)\!\sum_{k=0}^\infty \frac{\varepsilon_k}{2}(-)^k c_{kn}^{-m} \psi_k^{mc}
~~ &n\!+\!m\text{ even} \\
\!P_{n+1}^m(0)\!\sum_{k=1}^\infty\! (-)^k s_{kn}^{-m} \psi_k^{ms}  &n\!+\!m\text{ odd.}
\end{dcases} \label{irr_spherical_toroidal} 
\end{align}

$c_{nk}^m, s_{nk}^m$ are rational numbers with many equivalent explicit forms and recurrence relations (see appendix \ref{sec coefs}). For now they may defined by the stable recurrence:
\begin{align}
\!\!\bigg(\!n-m+\frac{1}{2}\!\bigg) c_{n+1,k}^m \!= (2k+1)c_{nk}^m + \bigg(\!n+m-\frac{1}{2}\!\bigg)c_{n-1,k}^m \label{c s rec}
\end{align}
and the same for $s_{nk}^m$. The initial values are
\begin{align}
c_{0k}^m&=\frac{(2m-1)!!}{2^{m-1}},\quad &c_{1k}^m&=-\frac{(2m-3)!!}{2^{m-1}}(2k+1), \nonumber\\
s_{0k}^m&=0, \quad &s_{1k}^m&=-\frac{(2m-3)!!}{2^m}(k+m+1).
\end{align}
(the double factorial can be extended to odd negative integers via recurrence, or equivalently through the gamma function). Note for $m<0$:
\begin{align}
c_{nk}^{-m}&=\frac{\Gamma(n-m+\frac{1}{2})}{\Gamma(n+m+\frac{1}{2})}c_{nk}^m, \nonumber\\
s_{nk}^{-m}&=\frac{\Gamma(n-m+\frac{1}{2})}{\Gamma(n+m+\frac{1}{2})}\frac{k-m+1}{k+m+1}s_{nk}^m.
\end{align}
The Legendre functions at 0 are, for $m\in\mathbb{Z}$:
\begin{align}
P_n^m(0)=\begin{dcases}
(-)^{(n-m)/2}\frac{(n+m-1)!!}{(n-m)!!} \quad &n+m\text{ even}\\ 0 &n+m\text{ odd}.
\end{dcases} 
\end{align}

\subsection{Existence of expansions} \label{sec existence}


Toroidal harmonics do not follow the same notion of internal and external as spherical and spheroidal harmonics. Ring harmonics $\Psi_m^{mv}$ can be written as a series of either internal or external spherical harmonics, due to the fact that they are finite at both the origin and at $r=\infty$. However the axial toroidal harmonics $\psi_n^{mv}$ are singular at the origin and  infinity, so cannot be expanded as a series of spherical harmonics at all \footnote{They can neither be expressed as a series of spherical harmonics of the second kind, $r^nQ_n(\cos\theta)$ or $r^{-n-1}Q_n(\cos\theta)$ which are also singular on the $z$ axis. This is due to differences in parity about $z$}. Also, neither internal and external spherical harmonics can be expressed as a series of ring harmonics $\Psi_n^{mv}$. This is because a series of ring harmonics can only converge outside some toroidal boundary, and this boundary must enclose the singularities of the function being expanded - the external spherical harmonics $S_n^m$ are singular at the origin, so this toroidal boundary must cover the origin and thus extend to all space, while the internal spherical harmonics $\hat{S}_n^m$ cannot be expanded for a similar reason - the torus must extend to all space to cover the ``singularity" at $r=\infty$. Contrarily, both internal and external spherical harmonics can be expanded with axial toroidal harmonics $\psi_n^{mv}$, because a series of axial toroidal harmonics will converge \textit{inside} some torus - for the internal spherical harmonics, this toroidal boundary may extend up to infinity since the functions are continuous in all space. For external spherical harmonics the toroidal boundary may extend to the origin. To check this we can determine the boundary of convergence of expansions \eqref{reg_spherical_toroidal} and \eqref{irr_spherical_toroidal} from the behaviour of the $k^{th}$ term in the series as $k\rightarrow\infty$. The Legendre functions grow as \cite{NIST:DLMF} pg 191 (\eqref{limP} is presented for completeness):
\begin{align}
\lim_{k\rightarrow\infty} P_{k-1/2}^m (\cosh\xi) =& \frac{k^m e^{k\xi}}{\sqrt{(2k-1)\sinh\xi}} \label{limP}\\
\lim_{k\rightarrow\infty} Q_{k-1/2}^m (\cosh\xi) =& \frac{\sqrt{\pi}(-k)^m e^{-k\xi}}{\sqrt{(2k-1)\sinh\xi}}  \label{limQ}
\end{align}
while $c_{kn}^{-m}$ and $s_{kn}^{-m}$ are bounded by the sequence $e_{k+1}=\frac{2n+1}{k} e_k + e_{k-1}$ (with the same initial values), which itself grows slower than $k^{2n+1}$. So the series coefficients decay geometrically as $e^{-k\xi}$, and $\xi>0$ everywhere except the $z$-axis and $r=\infty$, so the series converge everywhere, although slowly near the $z$ axis and for large $r$. In these cases the series terms grow significantly in magnitude before converging, which sacrifices accuracy because the series can only be accurate to the last digit of the largest term in the series. This causes problems in dealing with extremely tight tori.

\section{$T$-matrix on a spherical harmonic basis} \label{sec spherical}
For any bounded scatterer, the scattered field is expandable on a basis of outgoing spherical harmonics that converge outside the circumscribing sphere, and by linearity of Laplace's equation, the expansion coefficients are linearly related to the coefficients of the incident field. Therefore the $T$-matrix for any bounded scatterer should exist on a basis of spherical harmonics. In fact for a torus, the incident and scattered fields may both be expanded on interior or exterior spherical harmonics (see sec. \ref{reg irr}). First we look at the standard case where the external field is expanded on regular spherical harmonics and the scattered field will be expanded on irregular harmonics. So the potentials are assumed to have the following expansions:
\begin{align}
V_e&=\sum_{m=-\infty}^\infty\sum_{n=|m|}^\infty A_n^m \hat{S}_n^m, \\
V_s&=\sum_{m=-\infty}^\infty\sum_{n=|m|}^\infty C_n^m S_n^m, \label{sph_pot_expns}
\end{align}
the sum over $n$ can be written in matrix notation so that
\begin{align}
V_e&=\sum_m \b A^T \hat{\b S}, \label{Vex}\\
V_s&=\sum_m \b C^T \b S,       \label{Vsc} \qquad \b C=\b T\b A.
\end{align}
where for example $\b A$, $\hat{\b S}$ are $1\times N$ ($N$ being the numerical truncation order) column vectors containing the elements $A_n^m$ or $\hat S_n^m$ for all $n\geq 0$, for a fixed $m$, and $\b T$ is the $T$-matrix. The vectors and matrices start their index from $n=0$ even though the spherical harmonics are zero for $n<m$; this is match the dimensionality of the toroidal matrices which start from $n=0$. 

It is most likely not possible to expand the internal potential $V_i$ as a series of spherical harmonics, following the logic of section \ref{sec existence}. We know $V_i$ may be represented as a series of toroidal harmonics $\psi_n^m$ which are singular on the $z$-axis, so it is a fair assumption that the analytic continuation of $V_i$ is singular on atleast the $z$-axis (although not proven - it may be possible that a series of functions that are singular in some region could cancel exactly resulting in a smooth function, but given the complex series coefficients, this seems unlikely). This singularity would prohibit expansions in terms of both external or internal spherical harmonics since \textit{every} spherical annulus would contain singular points. This means that the $P$ and $Q$ matrices have no representation on a spherical basis, and therefore it would be impossible to compute the $T$-matrix directly via the extended boundary method on a spherical basis. However, since the problem can be solved using toroidal harmonics, we can use this and apply the spherical-toroidal transformations of section \ref{sec rels} without considering $V_i$.

For this it will be useful to express the relationships between spherical and toroidal harmonics in matrix notation:
\begin{align}
\hat{\b S} &= \b M^c_r\bm\uppsi^c + \b M^s_r\bm\uppsi^s, \label{phipsi}\\
\b\Psi^c &= \b N^c_i\b S, \qquad
\b\Psi^s = \b N^s_i\b S, \label{psiphi}
\end{align}
where the elements of matrices $\b M$ and $\b N$ can be determined from (\ref{cos_toroidal_spherical},\ref{sin_toroidal_spherical},\ref{reg_spherical_toroidal}) to be
\begin{align}
[M^c_r]_{nk}^m&=\frac{\varepsilon_k}{2\pi}P_n^m(0)c_{kn}^{-m}, \nonumber\\
[M^s_r]_{nk}^m&=\frac{-1}{\pi}P_{n+1}^m(0)s_{kn}^{-m}, \nonumber\\
[N^c_i]_{nk}^m&= P_k^{-m}(0)c_{nk}^m, \nonumber\\
[N^s_i]_{nk}^m&=-P_{k+1}^{-m}(0)s_{nk}^m. \label{MN}
\end{align}
The subscripts $r,i$ stand for regular or irregular (referring to the type of spherical harmonic), and the superscripts $c,s$ refer to cosine or sine of $\eta$. \\
First the excitation potential in \eqref{Vex} must be expressed on a basis of toroidal harmonics by applying \eqref{phipsi}:
\begin{align}
V_e=\sum_m \b A^T [ \b M^c_r\b\uppsi^c + \b M^s_r\bm\uppsi^s ]
\end{align}
then $V_s$ is found via the toroidal-basis $T$-matrix solution \eqref{Ttoroidal}:
\begin{align}
V_s=\sum_m \b A^T [ \b M^c_r (\bar{\b T}^c)^T \bm\Psi^c  + \b M^s_r (\bar{\b T}^s)^T \bm\Psi^s ]
\end{align}
and expanding the toroidal functions back into spherical harmonics with \eqref{psiphi}:
\begin{align}
V_s=\sum_m \b A^T [ \b M^c_r (\bar{\b T}^c)^T \b N^c_i + \b M^s_r (\bar{\b T}^s)^T \b N^s_i ] \b S.
\end{align}
Comparing this to \eqref{Vsc}, and noting that $\b C^T = \b A^T\b T^T$, the $T$-matrix is found to be
\begin{align}
\b T 
&= (\b N^c_i)^T \bar{\b T}^c (\b M^c_r)^T  + (\b N^s_i)^T \bar{\b T}^s (\b M^s_r)^T.
\end{align}

The $v=c$ term $(\b N^c_i)^T \bar{\b T}^c (\b M^c_r)^T$ contributes to the entries of $\b T$ with only $n$ and $k$ both even, while $(\b N^s_i)^T \bar{\b T}^s (\b M^s_r)^T$  contributes only to entries with $n$ and $k$ both odd. Hence $T_{nk}^m=0$ for $n+k$ odd as expected for particles with reflection symmetry about $z$.

Here we have not used normalised spherical basis functions, and as a consequence $\b T$ is not symmetric for $m>0$. The quasistatic limit of the conventional symmetric electromagnetic $T$-matrix in \cite{Mishchenko2002} can be obtained from:
\begin{align}
\lim_{k_1\rightarrow0}& T_{nk}^{22,m} = \frac{-i(k_1a)^{n+k+1}}{(2n-1)!!(2k-1)!!}\nonumber\\
&\times\sqrt{\frac{(n+1)(k+1)}{nk(2n+1)(2k+1)}\frac{(n+m)!(k-m)!}{(n-m)!(k+m)!}} T_{nk}^m
\end{align}
where $k_1$ is the wavenumber in the surrounding medium.

\subsection{Interior/exterior $T$-matrices} \label{reg irr}
In the above derivation we assumed that the external potential was created by a source that could be expanded as a series of \textit{regular} spherical harmonics, but because the torus excludes the origin, the external field may come from near the origin (for example a point source at the origin) and \textit{irregular} harmonics must be used instead. Similarly, to compute the scattered field near the origin, a regular basis must be used. 
The $T$-matrix derived above applies only if $V_e$ is expanded on $\hat S_m^m$ and $V_s$ is expanded on $S_n^m$, and we denote this matrix $\b T(r\rightarrow i)$. Following similar derivations, expressions for the other matrices can be found, and all four variations of these $T$-matrices are
\begin{align}
\b T(r\rightarrow i)&=(\b N^c_i)^T \bar{\b T}^c (\b M^c_r)^T  + (\b N^s_i)^T \bar{\b T}^s (\b M^s_r)^T \nonumber\\
\b T(r\rightarrow r)&=(\b N^c_r)^T \bar{\b T}^c (\b M^c_r)^T  + (\b N^s_r)^T \bar{\b T}^s (\b M^s_r)^T \nonumber\\
\b T(i\rightarrow r)&=(\b N^c_r)^T \bar{\b T}^c (\b M^c_i)^T  + (\b N^s_r)^T \bar{\b T}^s (\b M^s_i)^T \nonumber\\
\b T(i\rightarrow i)&=(\b N^c_i)^T \bar{\b T}^c (\b M^c_i)^T  + (\b N^s_i)^T \bar{\b T}^s (\b M^s_i)^T \label{Tmats}
\end{align}
The transformation matrices not already defined in \eqref{MN} differ only in alternating signs, and are
\begin{align}
[M^c_i]_{nk}^m&=\frac{\varepsilon_k(-)^k}{2\pi}P_n^m(0)c_{kn}^{-m},\nonumber\\
[M^s_i]_{nk}^m&=\frac{(-)^k}{\pi}P_{n+1}^m(0)s_{kn}^{-m}, 
\nonumber\\
[N^c_r]_{nk}^m&=(-)^nP_k^{-m}(0)c_{nk}^m, 
\nonumber\\
[N^s_r]_{nk}^m&=(-)^nP_{k+1}^{-m}(0)s_{nk}^m.
\end{align}
For example, $\b T(r\rightarrow r)$ applies if the source is outside the torus' circumscribing sphere, and if the scattered field is to be evaluated near the hole. 

For a conducting torus two pairs of these $T$-matrices in \eqref{Tmats} are identical. Mathematically it is straightforward to show that $\b T(i\rightarrow r)=\b T(r\rightarrow i)$. Physically this is due to the fact that the problem for a conductor does not distinguish the scattered and external/incident fields - the problem of determining the incident field from the scattered is the same as determining the scattered from the incident. And $\b T(r\rightarrow r)=\b T(i\rightarrow i)$, which can be shown by applying radial inversion about the sphere of radius $a$ centred at the origin. The geometry transforms as $r\rightarrow a^2/r$ which preserves the torus, while potentials transform as $V\rightarrow (a/r) V(r\rightarrow a^2/r)$, so $\hat{S}_n^m \leftrightarrow S_n^m$. This argument only applies to the conducting torus since permittivity also transforms as $\epsilon\rightarrow (a^2/r^2)\epsilon$.

From the static $T$-matrix we can make deductions of the applicability of the full-wave $T$-matrix. Since the high order limit of the spherical wave functions tends towards the solid spherical harmonics, the boundaries of convergence of the full wave solution are identical to that for the quasistatic solution. It is not possible to compute the scattered field in a spherical annulus around the mid section of the torus $r\sim a$, due to the singularity of the scattered field. Even in the case of a charged conducting torus (constant external potential), it can be shown that both  the interior and exterior spherical harmonic solutions diverge in the annulus $a<r<R_0$. This is analogous to spheroids with high aspect ratio - there is a region near the spheroid where the fields cannot be expressed using spherical basis functions.


\section{Derived physical quantities} \label{sec derived}
We now use these methods to calculate spatial fields, capacitance, polarizability, resonance conditions and cross sections.
 
\subsection{Spatial fields}
\begin{figure}
\includegraphics[scale=.7]{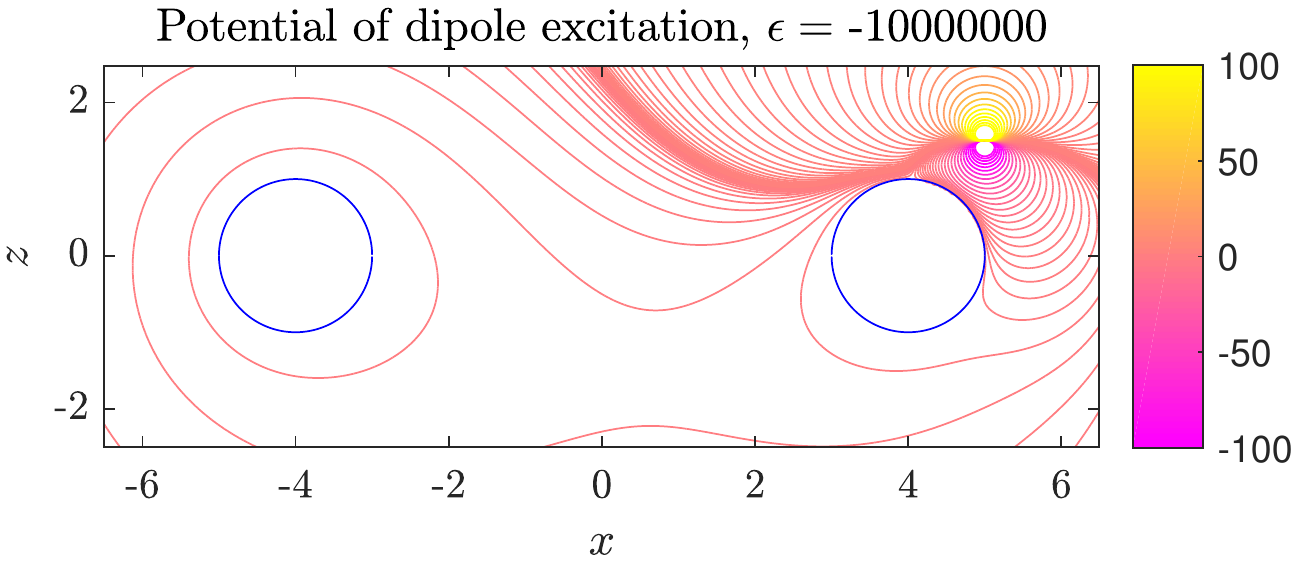}
\includegraphics[scale=.7]{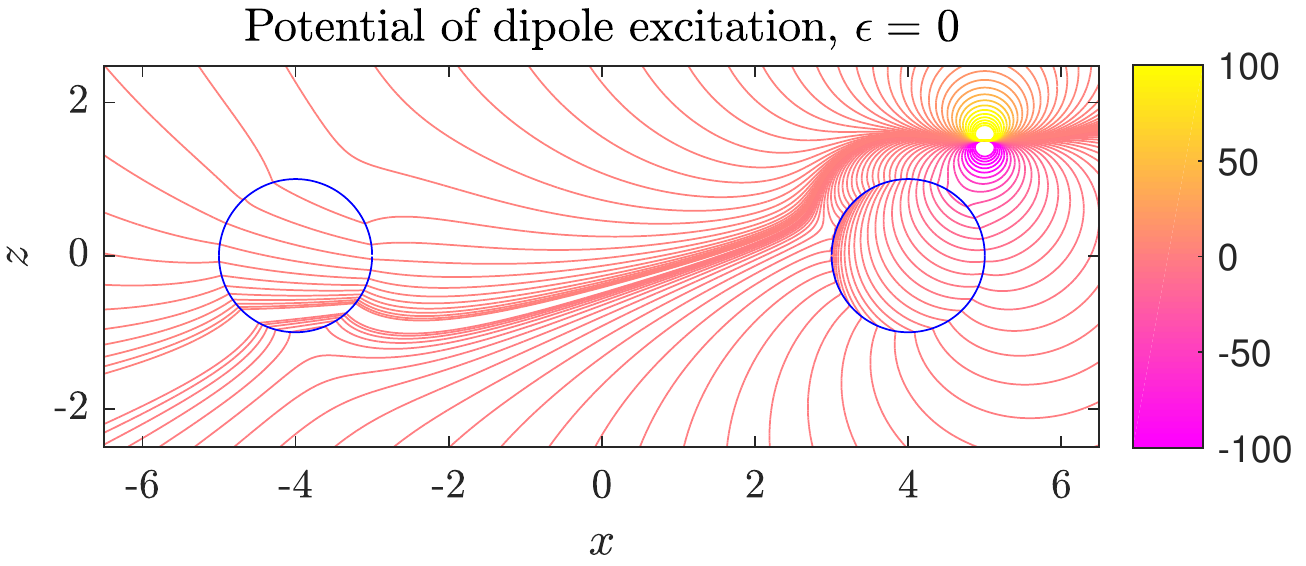}
\caption{Cross section of the equipotentials for a dipole located at $x=5,z=1.5$ for a torus with $R_0=4,r_0=1$, $\epsilon=0$ or $\epsilon=-10^7$. The contour lines are more densely spaced around small values of $V$ for better visualisation of the interaction with the torus (although this results in a dense accumulation around $V=0$). The potential was computed using toroidal T-matrices and toroidal harmonics as per supplemental codes attached. Matrices were truncated at $N=30$ for $\epsilon=-10^7$ and $N=50$ (approximately 15 s computation time) for $\epsilon=0$. The summation limit of $m$ is also taken to be $N$.} \label{potentialContour}
\end{figure}

To verify the boundary conditions for the potential we look at the extreme cases of a conducting ($\epsilon\rightarrow-\infty$) and extremely non-conducting torus ($\epsilon=0$) in figure \ref{potentialContour}. In both cases the equipotentials visually satisfy the boundary conditions in \eqref{BCs}. For a conducting torus the results for $\epsilon=-10^7$ coincide with conducting solution \eqref{conducting solution}. 

\begin{figure}
\includegraphics[scale=.7]{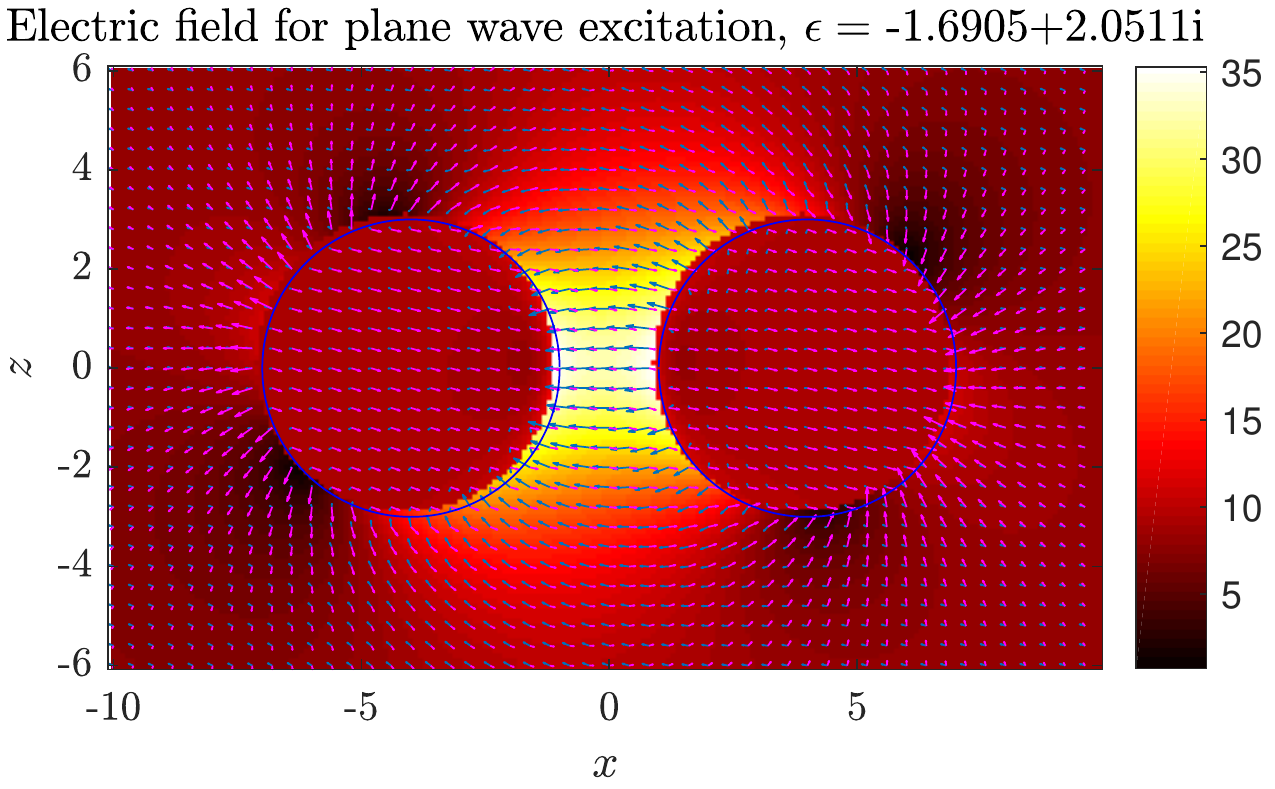}
\caption{Magnitude and field lines of the real part (blue) and imaginary part (magenta) of the electric field for a low frequency plane wave incident at an angle of $45\deg$ on a gold nano torus  in water. This is in the quasistatic limit, so the plane wave is effectively a uniform field but the permittivity of gold is taken at a wavelength of 500nm. The torus size is simply assumed to be much smaller than the wavelength ($\lesssim50nm$ across), so the scale and colourbar limits are arbitrary. The field was calculated via toroidal harmonics where the series and matrices were truncated at $N=20$ (about ~5s computation time). Circular polarization is seen where the real and imaginary vectors are not parallel.} \label{Efield}
\end{figure}

Since gold nanoparticles are often used for their field enhancements, the electric field is plotted for a gold nano-torus in water excited by a uniform field, showing a strong field enhancement in the gap. The plots in figures \ref{potentialContour}, \ref{Efield} have also been calculated using the spherical $T$-matrix and spherical harmonics, to verify that both approaches converge to the same result (atleast where the spherical solution converges).

\subsection{Capacitance and dipolar response of conducting torus}
The capacity $C$ of a conducting torus held at a uniform potential $V_0$ can be deduced from the induced charge $Q$, and is related to $T_{00}^0$:
\begin{align}
C&=4\pi\varepsilon_0aT_{00}^0, \\
T_{00}^0&=-\frac{2}{\pi}\sum_{q=0}^\infty\varepsilon_q\frac{Q_{q-1/2}(\beta_0)}{P_{q-1/2}(\beta_0)}. 
\end{align}
And for the dipole polarizability $\b \alpha$ we consider a uniform field along a cartesian axis, exciting dipole moment $\b p$ in a conducting torus. The polarizability is related to $T_{11}^0, T_{11}^1$:
\begin{align}
p_w&=\alpha_{ww}E_w,\qquad w=x,y,z,\\
\alpha_{zz}&=4\pi\epsilon_0 a^3 T_{11}^0, \\
\alpha_{xx}=\alpha_{yy}&=4\pi\epsilon_0 a^3 T_{11}^1,
\end{align}
where 
\begin{align}
T_{11}^0&=-\frac{16}{\pi}\sum_{q=1}^\infty q^2\frac{Q_{q-1/2}(\beta_0)}{P_{q-1/2}(\beta_0)}, \\
T_{11}^1&=\frac{1}{\pi}\sum_{q=0}^\infty\varepsilon_q (4q^2-1)\frac{Q_{q-1/2}^1(\beta_0)}{P_{q-1/2}^1(\beta_0)}.
\end{align}
These results agree with \cite{belevitch1983torus}.

\subsection{Plasmon resonances} \label{sec resonances}
\begin{figure}
\includegraphics[scale=.28]{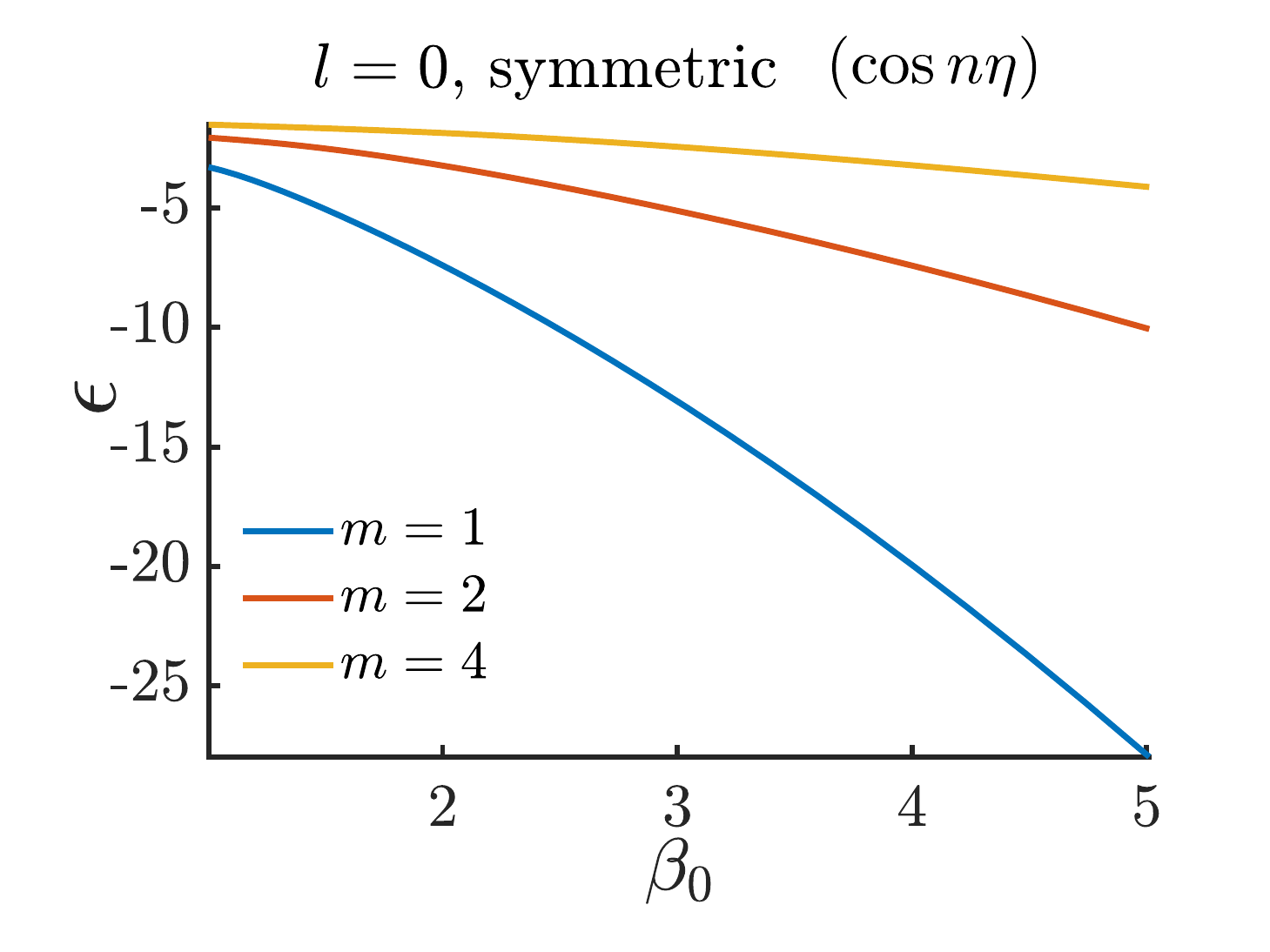}
\includegraphics[scale=.30]{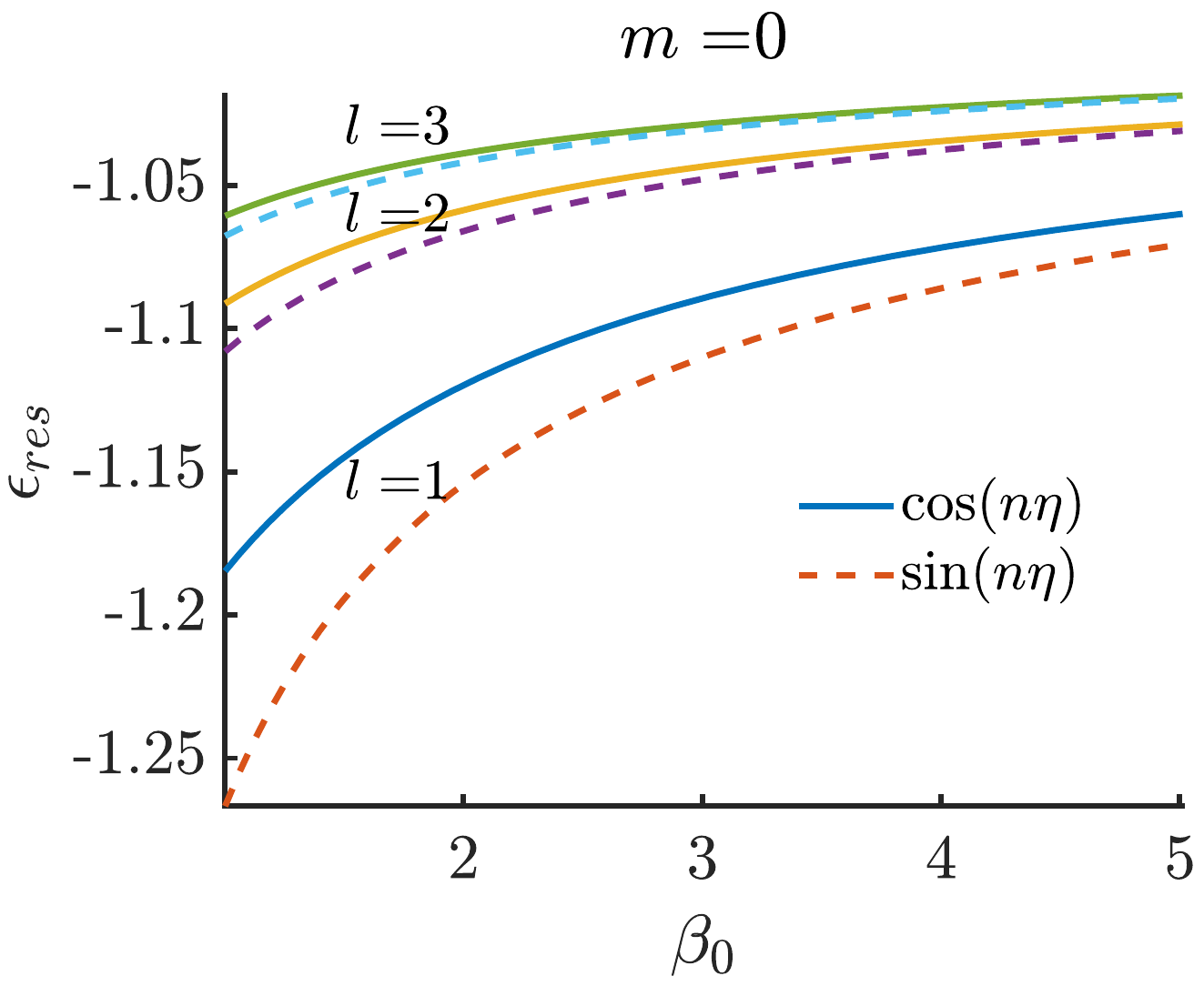}\\
\vspace{1mm}
\includegraphics[scale=.31]{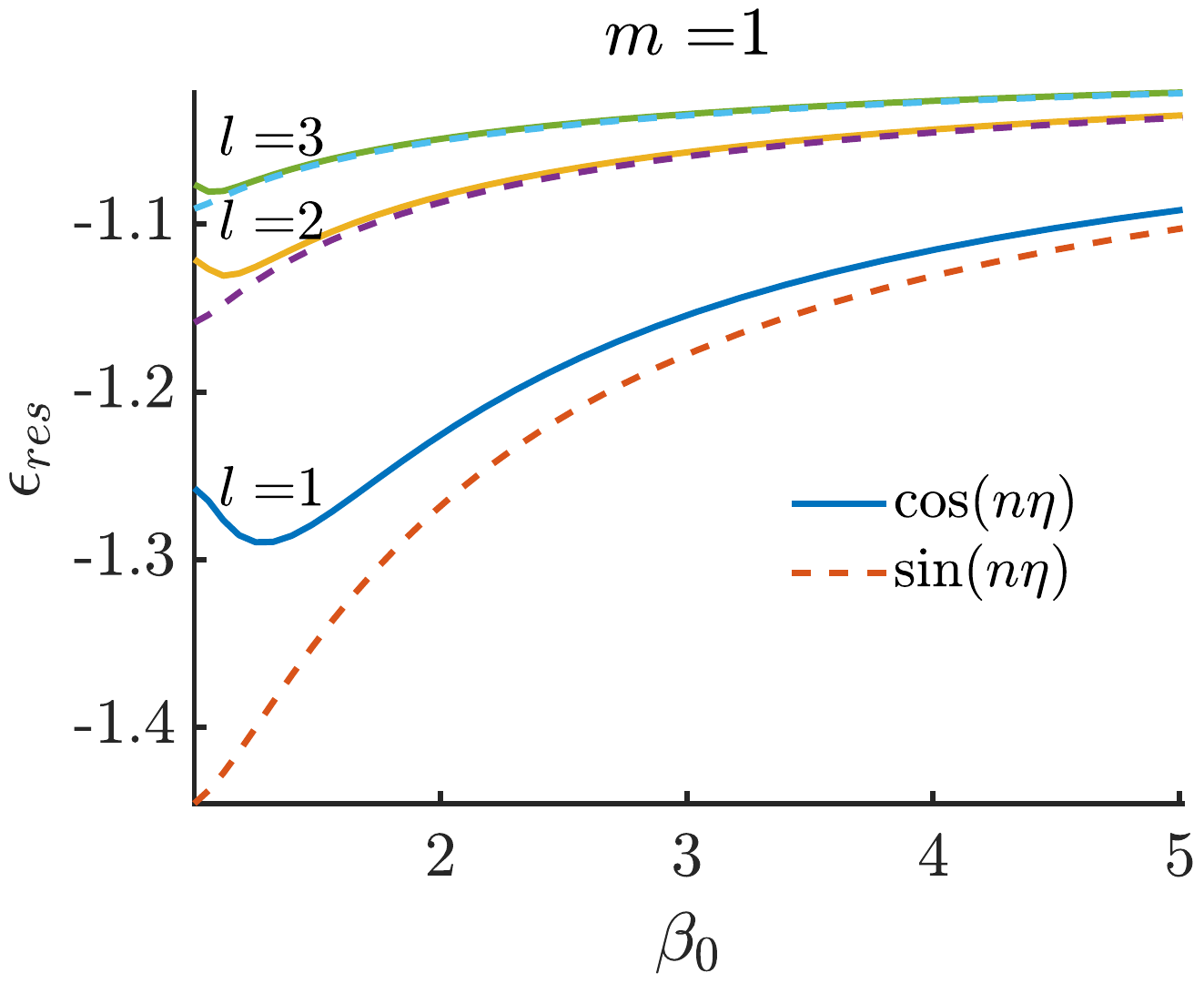}
\includegraphics[scale=.31]{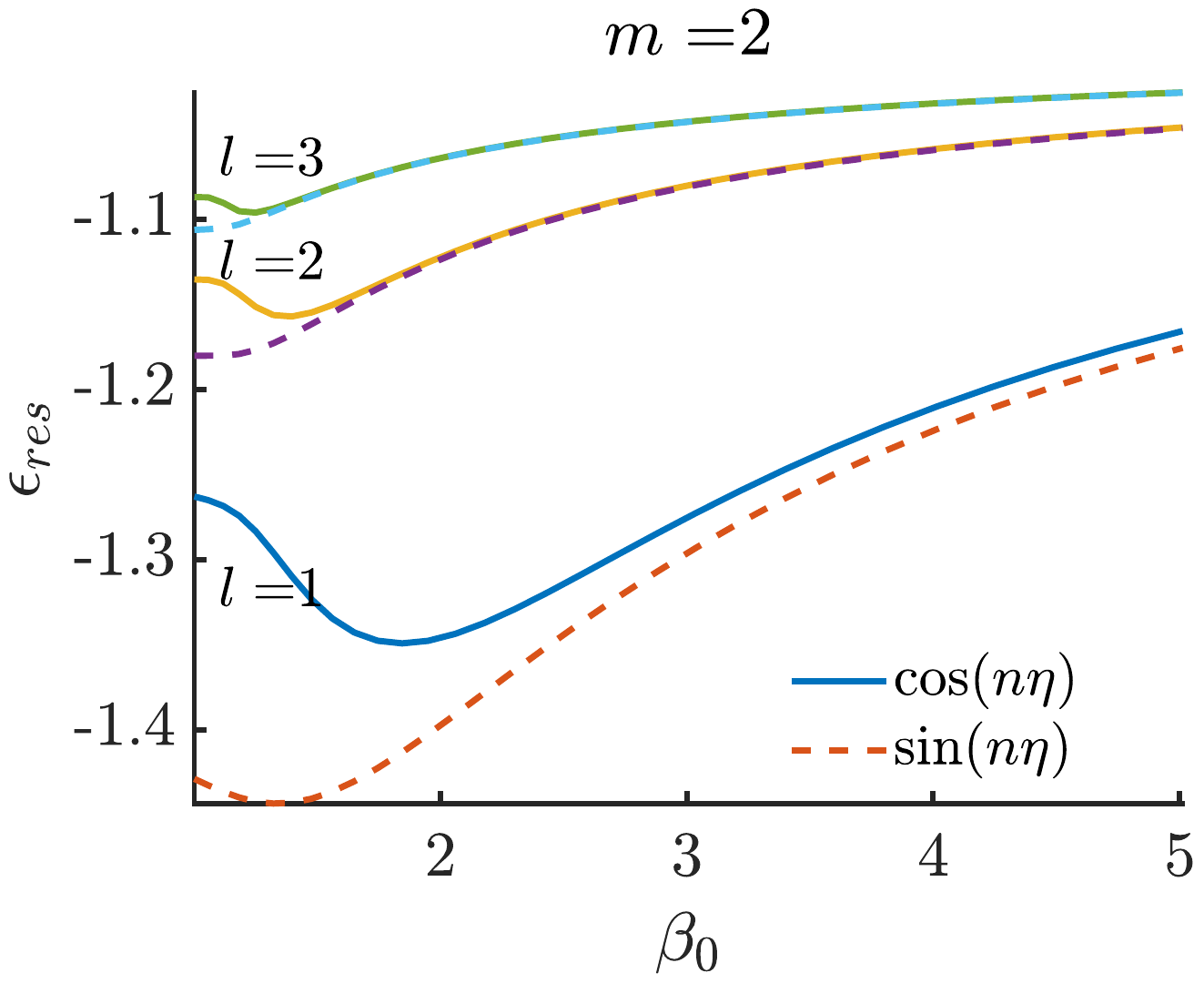}
\caption{Relative resonant permittivities of tori of aspect ratio $\beta_0=R_0/r_0$. The dashed (solid) curves correspond to the resonances of $\bar{\b T}^c$ ($\bar{\b T}^s$), which are symmetric (antisymmetric) about the torus plane. Resonances are labelled with mode number $l$ in order of increasing $\epsilon$. The antisymmetric $s$ modes start at $l=1$. The top left plot shows the strong resonances that only occur for $\bar{\b T}^c$. The other plots show the next three strongest resonances for both $v=c,s$; these resonances tend towards $\epsilon=-1$, for all values of $m,v,\beta_0$. The matrix size used in calculations ranged from $N=250$ for $\beta_0=1.01$ to $N=12$ for $\beta_0=5$.}
\label{resonances}
\end{figure}
From the explicit expressions for the toroidal matrices in (\ref{Qc},\ref{Qs}), we can calculate the values of $\epsilon$ that produce $V=\infty$ - a plasmon resonance. These are $\epsilon=\epsilon_l^{mv}$ for $l=0,1,2...$, $m=0,1,2...$, $v=c,s$. We have replaced the index $n$ with $l$ because each resonant mode consists of all toroidal harmonics for $n\geq0$, not accentuating one $n$ in particular. A natural choice is to let $l$ order these in terms of magnitude $|\epsilon_l^{mv}|$ ($\epsilon_l^{mv}$ are all negative real numbers), which appears to also order the resonances in terms of strength. $\epsilon_l^{mv}$ can be found from the condition $\det(\bar{\b Q})=0$, through $\epsilon_l^{mv}=1-1/\lambda_l^{mv}$ where $\lambda_l^{mv}$ are the eigenvalues of $\hat{\b Q}^{mv}$ and $\bar{\b Q}^{mv}=\b I-(\epsilon-1)\hat{\b Q}^{mv}$. Note that $\hat{\b Q}$ is independent of $\epsilon$.

In figure \ref{resonances} the conditions for resonances are plotted as a function of aspect ratio $\beta_0$, and agree with the continued fraction approach  \cite{avramov1993high,love1975quasi,garapati2017poloidal}. Also, these results confirm the discussion of \cite{salhi2016thesis} claiming that $m=0$ resonances exist, even though they were not obtained in \cite{love1975quasi,avramov1993high}. 

\subsection{Optical cross sections} 
\begin{figure}
\includegraphics[scale=.61]{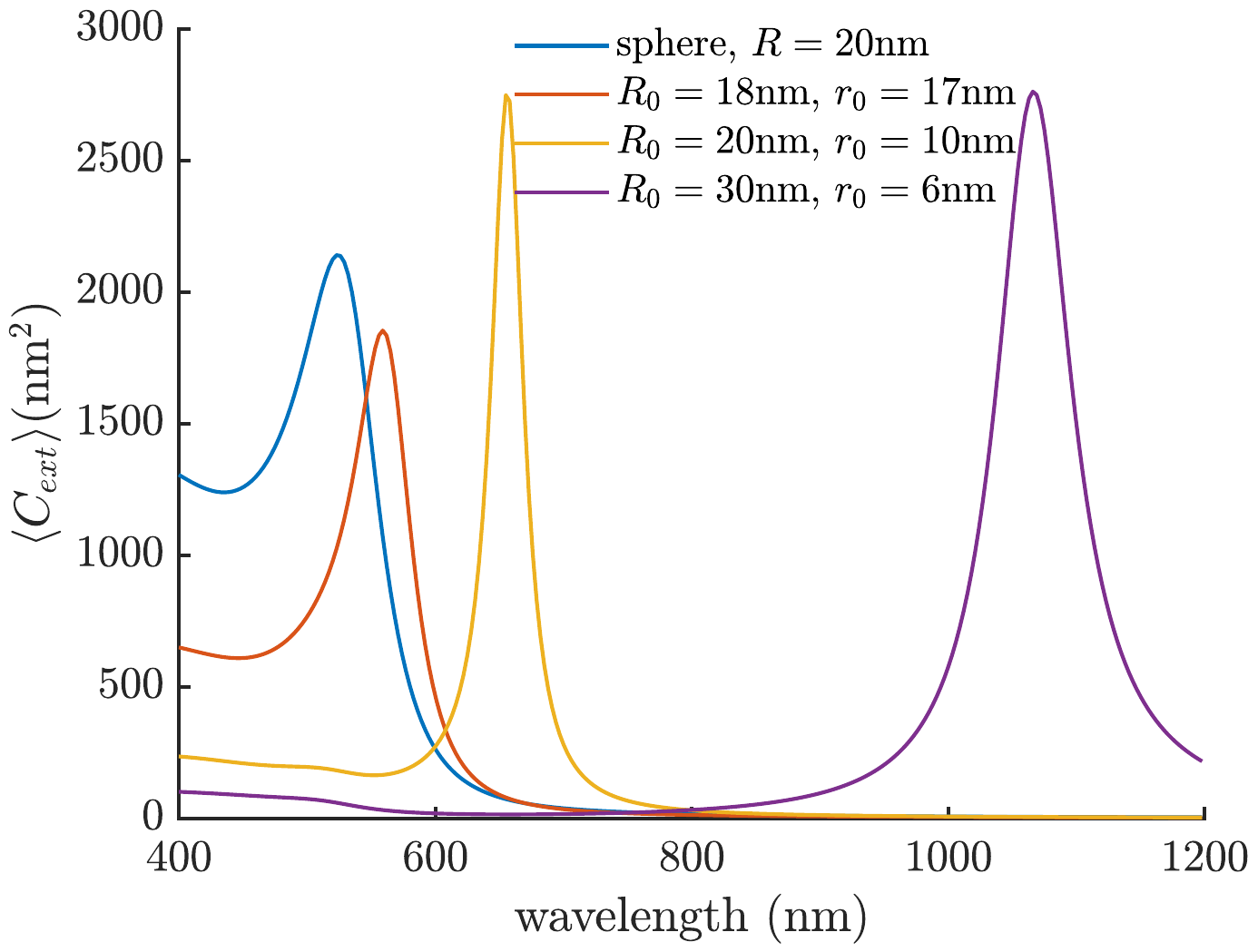}
\caption{Extinction cross sections for a sphere and gold nano-tori in water. The dielectric function of gold was taken from eq. (E.2) of \cite{LeRu2009}, and $\epsilon_\text{water}=1.77$. The radiative correction has been applied.} \label{CextTorus} 
\end{figure}

The cross sections are obtained from the elements of the full-wave $T$-matrix \cite{Mishchenko2002}. In the small particle limit, only the dominant terms $T^{22,m}_{11}$ for $m=0,1$ contribute. For example we look at the orientation averaged cross-sections:
\begin{align}
\lim_{k_1\rightarrow0}\langle C_{ext}\rangle&=-\frac{2\pi}{k_1^2}\text{Re}\{T_{11}^{22,0}+2T_{11}^{22,1}\}, \\ 
\lim_{k_1\rightarrow0}\langle C_{sca}\rangle&=\frac{2\pi}{k_1^2}\big(|T_{11}^{22,0}|^2+2|T_{11}^{22,1}|^2\big),  
\end{align}
with
\begin{align}
\lim_{k_1\rightarrow0}T_{11}^{22,0}&= -i(k_1a)^3\frac{32}{3\pi}\sum_{n=0}^\infty\sum_{k=0}^\infty nk~ \bar{T}_{nk}^{m=0,s}, \label{T110}\\
\lim_{k_1\rightarrow0}T_{11}^{22,1}&= ~~i(k_1a)^3\frac{2}{3\pi}\sum_{n=0}^\infty\sum_{k=0}^\infty \varepsilon_k (4n^2-1)\bar{T}_{nk}^{m=1,c}. \label{T111}
\end{align}
The extinction cross-section is plotted in figure \ref{CextTorus} for various nano-tori and appear consistent with results from \cite{el2009torus,salhi2015toroidal,mary2007optical}. The dominant response comes from $T^{22,1}_{11}$ - excitation along the $xy$- plane - this is the $l=0,m=1$ plasmon resonance in figure \ref{resonances}. We cannot see the resonances for $l\geq1$ that occur for small negative epsilon due to the considerable imaginary part of the dielectric function of gold. On the other hand, $T^{22,0}_{11}$ is almost negligible and has no resonance.
 
\section{Thin ring limit}
\label{sec thin}

We can find analytic results for the matrix elements in the thin ring limit. The limits of the Legendre functions are \cite{NIST:DLMF}
\begin{align}
P_{-1/2}^m(\beta_0)&\rightarrow \frac{\sqrt{2/(\pi\beta_0)}}{\Gamma(-m+\frac{1}{2})}\bigg[\log(8\beta_0)-2\sum_{p=1}^{m}\frac{1}{2n-1}\bigg] \nonumber\\
&\hspace{4cm} m\geq0 \label{P-1/2} \\
P_{n-1/2}^m(\beta_0)&\rightarrow \frac{(n-1)!(2\beta_0)^{n-1/2}}{\sqrt{\pi}\Gamma(n-m+\frac{1}{2})}, \quad\qquad n>0 \\
Q_{n-1/2}^m(\beta_0)&\rightarrow \frac{(-)^m\sqrt{\pi}\Gamma(n+m+\frac{1}{2})}{n!(2\beta_0)^{n+1/2}},
\end{align} 
and we will need $Q_{-1/2}^m(\beta_0)$ to second order:
\begin{align}
Q_{-1/2}^m(\beta_0)&\rightarrow \frac{\pi(-)^m(2m-1)!!}{2^m\sqrt{2\beta_0}}\bigg[1+\frac{4m^2+3}{16\beta_0^2}\bigg].
\end{align}
which can be obtained from its hypergeometric function definition.
\eqref{P-1/2} agrees with \cite{Scharstein2005} as $\beta\rightarrow\infty$ for $m=0,1,2,...$, but is more accurate.
It can be shown that the toroidal $Q$-matrices in this limit are lower triangular - $\bar{Q}_{nk}^{mc}=\bar{Q}_{nk}^{ms}=0$ for $n<k$, with
\begin{align}
\bar{Q}_{nk}^{m\substack{c\\s}}&\rightarrow \delta_{nk}\frac{\epsilon+1}{2} + (1-\delta_{nk})(1\pm\delta_{k0})\nonumber\\
&\times\frac{\epsilon-1}{4}\frac{\Gamma(k+m+\frac{1}{2})}{\Gamma(n+m+\frac{1}{2})}\frac{(n-1)!}{k!}  \qquad \quad n\geq k, n\neq0\nonumber\\
\bar{Q}_{00}^{mc}&\rightarrow 1 + \frac{\epsilon-1}{\beta_0^2}\frac{4m^2+1}{8}\bigg[\log(8\beta)-2\sum_{p=1}^{|m|}\frac{1}{2p-1}\bigg]\nonumber\\
\bar{Q}_{00}^{ms}&= 1 \quad \forall \beta_0. \label{Q thin}
\end{align}
Now we use the $R$-matrix $\bar{\b R}=\bar{\b Q}^{-1}$, which has the following limit:
\begin{align}
\bar{R}_{nk}^{m,\substack{c\\s}}\rightarrow& \frac{2\delta_{nk}}{\epsilon+1} +(1-\delta_{nk})\bigg( {1+\epsilon\delta_{k0}\atop1-\delta_{k0}}\bigg)\frac{\epsilon-1}{(\epsilon+1)^{n-k+1}}\nonumber\\
&\times\frac{\Gamma(k+m+\frac{1}{2})}{\Gamma(n+m+\frac{1}{2})}\prod_{p=k+1}^{n-1}\bigg(\frac{\epsilon-1}{2}+p(\epsilon+1)\bigg), \nonumber\\ &\hspace{5cm}\quad n\geq k, n\neq0 \label{R thin}
\end{align}
This expression for $\bar{\b R}$ is the inverse of $\bar{\b Q}$ in \eqref{Q thin}, i.e. $\sum_{p=0}^\infty \bar{Q}_{np} \bar{R}_{pk}=\delta_{nk}$, which can be proven with the help of Mathematica. To obtain a result for the $T$-matrix entry $T_{00}^m$, it is also necessary to include the second order for the element $R_{00}^{mc}$:
\begin{align}
\bar{R}_{00}^{mc}&\rightarrow 1 - \frac{\epsilon-1}{\epsilon+1}\frac{2m^2(\epsilon\!+\!1)\!+\!1}{4\beta_0^2}\bigg[\!\log(8\beta)-2\sum_{p=1}^{|m|}\!\frac{1}{2n-1}\bigg]\nonumber\\
\bar{R}_{00}^{ms}&= 1 \quad \forall \beta_0.
\end{align}
The second order for $\bar{R}_{00}^{mc}$ is obtained by including the 2nd order element $Q_{01}^{mc}$ (above the diagonal) and inverting the matrix  consisting of just the top left $2\times2$ block of $\bar{\b Q}^{mc}$. Then the $T$-matrix \eqref{Tthin} is obtained through \eqref{ThatP}:
\begin{align}
\bar{T}_{nk}^{mv}\rightarrow& (-)^m\frac{\pi\Gamma(n+m+\frac{1}{2})\Gamma(n-m+\frac{1}{2})}{n!(n-1)!(2\beta_0)^{2n}}(\bar{R}_{nk}^{mv}-\delta_{nk})\nonumber\\
&\hspace{4cm} ~~~ n\geq k, n\neq0 \nonumber\\
\bar{T}_{00}^{mc}\rightarrow& -\frac{\pi^2}{8}\frac{\epsilon-1}{\epsilon+1}\frac{2m^2(\epsilon+1)+1}{\beta_0^2}\nonumber\\
\bar{T}_{n0}^{ms}=&~\bar{T}_{0k}^{ms}=0.\label{Tthin} 
\end{align}

The upper diagonal entries of $\bar{\b T}^{mv}$ with $n\leq k$ can be obtained from the symmetry property of the toroidal $T$-matrix \eqref{symmetry}. In this limit there is just one resonance at $\epsilon=-1$, consistent with the trends seen in figure \ref{resonances} for large $\beta_0$, and the resonance for a cylinder. All elements have been checked numerically against the exact matrix elements in section \ref{sec toroidalTmat}.\\

An important feature is that the top left 2$\times$2 square of $\bar{\b T}^{mc}$ are all order $\beta_0^{-2}$, while all other entries decay quickly as $\beta_0\rightarrow\infty$, meaning that modes for $n= 0, 1$ dominate both the excitation ($V_e$) and response ($V_s$). For $\bar{\b T}^{ms}$, there is only one dominant element, $T_{11}^{ms}$.\\

We can now determine limits for the static dipolar polarizabilities per unit volume, to $\mathcal{O}(\beta_0^{-2})$, using (\ref{T110},\ref{T111}) with (\ref{Tthin}):
\begin{align}
\alpha_{zz}^\text{ring}&\rightarrow \label{alphazz} 2\frac{\epsilon-1}{\epsilon+1},\\
\alpha_{xx}^\text{ring}&\rightarrow \frac{\epsilon+3}{2}\frac{\epsilon-1}{\epsilon+1}. 
\label{alphaxx}
\end{align} 
These produce reasonable accuracy ($\lesssim10\%$) for aspect ratios $\beta_0\gtrsim5$, but fail when $|\epsilon|>>\beta_0$. 
It is interesting to compare these to the polarizabilities of a thin prolate spheroid, or needle:
\begin{align}
\alpha_{zz}^\text{needle}&\rightarrow \epsilon-1,\\
\alpha_{xx}^\text{needle}&\rightarrow 2\frac{\epsilon-1}{\epsilon+1}. 
\end{align}
$\alpha_{zz}^\text{ring}$, $\alpha_{xx}^\text{needle}$ apply when the long dimension of the wire is perpendicular to the applied electric field, and in fact both are of the form $\alpha_\perp=2(\epsilon-1)/(\epsilon+1)$. On the other hand $\alpha_{xx}^\text{ring}$ and $\alpha_{zz}^\text{needle}$ both diverge as $\epsilon\rightarrow\infty$, where the long dimension of the wire is aligned with the electric field. $\alpha_{zz}^\text{needle}$ is of the form $\alpha_{||}=\epsilon-1$. Intuitively, for $\alpha_{xx}^\text{ring}$ the ring is half aligned perpendicular and half parallel to the incident field, so that $\alpha_{xx}^\text{ring}=(\alpha_\perp+\alpha_{||})/2$. \\

Numerical tests show that for very large |$\epsilon$| and very thin rings, the approximate expressions for $\bar{\b R}^{m,c}$ break down, particularly for $m=0$. For analysis of the thin ring limit for conducting tori, see \cite{Scharstein2005}.

\section{Conclusion}
The problem of a dielectric torus in an arbitrary electrostatic field has been solved semi-analytically, and expressions are found for the $T$-matrix on both a toroidal and spherical harmonic basis. New forms of the series relationships between spherical and toroidal harmonics were derived. Resonant permittivities are calculated from matrix eigenvalues and agree with results from solving a continued fraction equation. Fully analytic asymptotic expressions are given in both the conducting and thin ring limits. The $T$-matrix has been linked the time-harmonic $T$-matrix governing electric multipole interactions, $\b T^{22}$. These results prove the existence of the $T$-matrix for such a complex shape, atleast in the small size limit. It is also found that similar $T$-matrices can be defined depending on whether the source is near or far from the torus hole and whether the scattered field is to be evaluated near or far from the origin. 

Hopefully this approach of applying basis transformations to the spherical harmonics and evaluating the surface integrals can be used to numerically obtain the $T$-matrix for other objects with ring like geometry.

To extend these results to the Helmholtz equation, we would need a basis suited to the toroidal geometry but unfortunately the Helmholtz equation is not separable in toroidal coordinates. A basis of toroidal wavefunctions were presented in \cite{weston1960toroidal}, although these are not orthogonal.

A similar work is in preparation on bispherical coordinates and scattering by two spheres. Bispherical and toroidal coordinates essentially differ by a real/imaginary focal distance, and the corresponding harmonics differ by a shift of the separation constant by $\pm1/2$, however, the topology of the two systems is radically different.

\acknowledgements{I wish to thank my supervisor Eric C. Le Ru for helpful discussions, and Victoria University of Wellington for financial support through a Victoria Doctoral Scholarship.}

\appendix

\section{Evaluation of the surface integral \eqref{Tmat int cos}} \label{int proof}
First we express the integral using the product formula:
\begin{align}
\int_{-\pi}^\pi \frac{\cos(n\eta)\cos(k\eta)}{\cosh\xi-\cos\eta}\d\eta &= \frac{1}{2}\int_{-\pi}^\pi \frac{\cos((n+k)\eta)}{\cosh\xi-\cos\eta}\d\eta  \nonumber\\ &+\frac{1}{2}\int_{-\pi}^\pi \frac{\cos((n-k)\eta)}{\cosh\xi-\cos\eta}\d\eta. \label{Tmat int app}
\end{align}
Following the method of \cite{mason2002chebyshev} (lemma 8.3) with minor changes we make the substitutions $z=e^{i\eta}$,  $p=n+k$, and rewrite the first integral  as
\begin{align}
\frac{1}{2}\int_{-\pi}^\pi &\frac{\cos(p\eta)}{\cosh\xi-\cos\eta}\d\eta \nonumber\\&= 
\int_{|z|=1}i\frac{z^p+z^{-p}}{\sinh\xi}\left(\frac{1}{z-e^\xi} - \frac{1}{z-e^{-\xi}}\right)\d z
\end{align}
There are two poles inside the integration path, a simple pole at $z=e^{-\xi}$ and a pole of order $p$ at $z=0$. Applying the residue theorem we find:
\begin{align}
\frac{1}{2}\int_{-\pi}^\pi \frac{\cos(p\eta)}{\cosh\xi-\cos\eta}\d\eta =\frac{\pi e^{-p\xi}}{\sinh\xi}. 
\end{align}
Then combine this with \eqref{Tmat int app} to obtain \eqref{Tmat int cos}. The proof of \eqref{Tmat int sin} is similar.

\section{Derivations of the relationships between spherical and toroidal harmonics}

\subsection{Expansions of toroidal harmonics} \label{toroidal vs spherical appendix}
We can first obtain the expansions of toroidal harmonics $\Psi_0^{mc}$ in terms of spherical harmonics $S_n^m, \hat{S}_n^m$, by equating two expansions of Green's function. For points $\b r_1$ and $\b r_2$ with $r_1<r_2$ , the spherical harmonic expansion of Green's function is \cite{Morse1953}
\begin{align}
\frac{1}{|\b r_1-\b r_2|}=&\sum_{m=-\infty}^\infty\sum_{n=|m|}^\infty(-)^m\frac{r_1^n}{r_2^{n+1}}P_n^m(u_1)P_n^{-m}(u_2)\nonumber\\
&\hspace{3cm}\times\cos m(\phi_1\!-\!\phi_2).  \label{GFP}
\end{align}
In this document $P_k^m(u)$ are defined without the phase $(-)^m$, while $P_n^{-m}=(-)^m\frac{(n-m)!}{(n+m)!}P_n^m$. \\
Also we have the 'cylindrical' expansion \cite{Cohl1999}:
\begin{align}
\frac{1}{|\b r_1-\b r_2|}=&\frac{1}{\pi\sqrt{\rho_1\rho_2}}\sum_{m=-\infty}^\infty Q_{m-1/2}(\chi_{12})\cos m(\phi_1-\phi_2), \nonumber\\ \chi_{12}=&\frac{\rho_1^2+\rho_2^2+(z_1-z_2)^2}{2\rho_1\rho_2} \label{GFQ}
\end{align}
which converges in all space except at $\b r_1\neq \b r_2$. Note $Q_{-m-1/2}^n=Q_{m-1/2}^n$.

Evaluating both \eqref{GFP} and \eqref{GFQ} at $\rho_2=a,z_2=0$ ($\Rightarrow u_2=0$, $r_2=a$, $\chi_{12}=\chi=\beta/\sqrt{\beta^2-1}$), and equating the $m^{th}$ term gives for $r<a$:
\begin{align}
\sqrt{\frac{a}{\rho}}
Q_{m-1/2}(\chi)=\pi\sum_{k=|m|}^\infty (-)^mP_k^{-m}(0)\left(\frac{r}{a}\right)^kP_k^m(u) \label{relation1r<a}
\end{align} 
Similarly the expansion for $r>a$ can be found by setting $\rho_1=a, z_1=0$ in \eqref{GFP} and \eqref{GFQ}:
\begin{align}
\sqrt{\frac{a}{\rho}}
Q_{m-1/2}(\chi)=\pi\sum_{k=|m|}^\infty (-)^mP_k^{-m}(0)\left(\frac{a}{r}\right)^{k+1}P_k^m(u) \label{relation1r>a}
\end{align}
The left hand side is actually proportional to the toroidal harmonic $\Psi_0^{mc}$. This can be seen by application of the Whipple formulae, which expressed in toroidal coordinates read as
\begin{align}
\Delta P_{n-1/2}^m(\beta)=\frac{(-)^n2/\sqrt{\pi}}{\Gamma(n-m+\frac{1}{2})}\sqrt{\frac{a}{\rho}}Q_{m-1/2}^n(\chi) \label{WhipP}
\\
\Delta Q_{n-1/2}^m(\beta)=\frac{(-)^n\pi\sqrt{\pi}}{\Gamma(n-m+\frac{1}{2})}\sqrt{\frac{a}{\rho}}P_{m-1/2}^n(\chi). 
\label{WhipQ}
\end{align}
We may then rewrite the expansions (\ref{relation1r<a},\ref{relation1r>a}) as 
\begin{align}
\Delta P_{-1/2}^m(\beta)=&
\frac{2\sqrt{\pi}(-)^m}{\Gamma(-m+1/2)}\sum_{k=|m|}^\infty P_k^{-m}(0)\nonumber\\&\times
\begin{dcases}
\left(\frac{r}{a}\right)^kP_k^m(u) \quad &r<a \\
\left(\frac{a}{r}\right)^{k+1}P_k^m(u) \quad &r>a  
\end{dcases}\label{relation1}
\end{align} 
which is \eqref{cos_toroidal_spherical} for $n=0$. 
\eqref{relation1} has been generalized to the Helmholtz equation (harmonic time dependence), as a spherical wave function expansion of a circular ring with current distribution expressed as a Fourier series \cite{Hamed2014}.\\

The $n=1$ toroidal harmonics $\Psi_1^{ms}$ are the potential of a ring of dipoles pointing in the $z$-direction. This can be obtained by applying the operator $\pd_z$ which transforms a charged ring into a double ring with dipole moment in the $z$-direction. Explicitly:
\begin{align}
a\frac{\pd}{\pd z}\Psi_0^{mc}=\left(m-\frac{1}{2}\right)\Psi_1^{ms}.
\end{align}
$\pd_z$ is also a ladder operator for the spherical harmonics:
\begin{align}
a\pd_z \hat{S}_n^m&=(n+m)\hat{S}_{n-1}^m \\
a\pd_z S_n^m&=-(n-m+1)S_{n+1}^m
\end{align}
Applying $a\pd_z$ to the toroidal harmonic expansion \eqref{relation1}, and noting an identity for the derivative of the Legendre functions, leads directly to the expansion for $\Psi_1^{ms}$.\\

$\Psi_1^{mc}$ is the potential of a ring of dipoles oriented outwards from the origin. To generate this we apply $r\pd_r$ which preserves harmonicity as does $\pd_z$, but turns a ring of charge on the $xy$ plane into a ring of dipoles pointing inward, plus, as it turns out, a net charge on the ring, as noted in section \ref{sec ring charge}. \\
Applying $r\pd_r$ to the $n=0$ toroidal harmonic expansion \eqref{relation1} and rearranging gives the expansion for $\psi_1^{mc}$.
We can now derive the expansions for general $n$ by repeated application of $r\pd_r$.
For the spherical harmonics:
\begin{align}
r\frac{\pd}{\pd r}S_n^m &= -(n+1)S_n^m \\
r\frac{\pd}{\pd r}\hat S_n^m &= n \hat S_n^m.
\end{align}
applying $r\pd_r$ to $\Psi_n^{mv}$ and rearranging, making use of the product to sum formulae for trigonometric functions and the recurrence relation for the Legendre functions $P_{n-1/2}^m$ for increasing $n$, we obtain:
\begin{align}
2r\frac{\pd}{\pd r}\Psi_n^{mv} &= \!\left(n+m-\frac{1}{2}\right)\Psi_{n-1}^{mv} - \Psi_n^{mv} \nonumber\\ &-\left(n-m+\frac{1}{2}\right) \Psi_{n+1}^{mv}.  \label{rdrQ}
\end{align}
By assuming some expansion coefficients $c_{nk}^m$, $s_{nk}^m$ for the expansions of the toroidal harmonics as in (\ref{cos_toroidal_spherical},\ref{sin_toroidal_spherical}), we can deduce that they satisfy the recurrence relation \eqref{c s rec}, which is sufficient to calculate the coefficients for all $n,k,m$ without numerical problems.

The first few orders for $m=0$ are:
\begin{align*}
c_{0,k}^0&=2 \\ 
c_{1,k}^0&=2(2k+1) \\ 
c_{2,k}^0&=\frac{16}{3} (k^2+k+\frac{3}{8}) \\ 
c_{3,k}^0&=\frac{32}{15}(2k+1)(k^2+k+\frac{15}{16}) \\ 
c_{4,k}^0&=\frac{64}{105}(4k^4+8k^3+15k^2+11k+\frac{105}{32}) \\
c_{5,k}^0&=\frac{128}{945}(2k+1)(4k^4+8k^3+\frac{109}{4}k^2+\frac{93}{4}k+\frac{945}{64}) \\
s_{0,k}^0&=0 \\
s_{1,k}^0&=4(k+1) \\
s_{2,k}^0&=\frac{8}{3}(k+1)(2k+1) \\
s_{3,k}^0&=\frac{64}{15}(k+1)(k^2+k+\frac{13}{16}) \\
s_{4,k}^0&=\frac{128}{105}(k+1)(2k+1)(k^2+k+\frac{76}{32}) \\
s_{5,k}^0&=\frac{256}{945}(k+1)(4k^4+8k^3+\frac{107}{4}k^2+\frac{91}{4} k+\frac{789}{64})
\end{align*}

Although these coefficients may be computed for all integer $n,k,m$, $c_{nk}^m$ are only relevant for $k+m$ even, while $s_{nk}^m$ only for $k+m$ odd.

\subsection{Expansions of spherical harmonics}
We start by applying a trigonometric identity to rearrange the toroidal Green's function expansion \cite{Morse1953} as 
\begin{align}
&\frac{1}{|\b r_1-\b r_2|}
=\frac{\Delta_1\Delta_2}{2\pi a}\sum_{m=-\infty}^\infty \sum_{k=-\infty}^\infty P_{k-1/2}^m(\beta_1)Q_{k-1/2}^{-m}(\beta_2)\nonumber\\
&\times[\cos(k\eta_1)\cos(k\eta_2)+\sin(k\eta_1)\sin(k\eta_2)]\cos m(\phi_1\!-\!\phi_2), \nonumber\\
&\hspace{5cm} \beta_1<\beta_2,\label{GF}
\end{align}
and substitute the spherical harmonic expansions for point $\b r_2$ (\ref{cos_toroidal_spherical},\ref{sin_toroidal_spherical}),
\begin{align}
&\frac{1}{|\b r_1-\b r_2|}=\frac{\Delta_2}{2\pi a}\sum_{m=-\infty}^\infty \sum_{k=-\infty}^\infty Q_{k-1/2}^{-m}(\beta_2)(-)^k 
\nonumber\\
&\times \sum_{n=|m|}^\infty\left[\cos(k\eta_2) c_{kn}^mP_n^{-m}(0)+ \sin(k\eta_2)s_{kn}^mP_{n+1}^{-m}(0)\right] \nonumber\\
&\times\left(\frac{r_1}{a}\right)^nP_n^m(u_1)\cos m(\phi_1-\phi_2),
\end{align}
and compare this to the spherical expansion of Green's function \eqref{GFP}, equating the coefficients of $P_n^m(u_1)$ for all $n,m$. This gives the expansion of spherical harmonics $S_n^m,\hat{S}_n^m$ in terms of toroidal harmonics $\psi_n^{mv}$, as presented in (\ref{reg_spherical_toroidal},\ref{irr_spherical_toroidal}).\\

Looking at the low orders, for $n=0$, \eqref{reg_spherical_toroidal} is the expansion of a constant onto the basis of toroidal harmonics, and is known as Heine's expansion. And $n=0$ for \eqref{irr_spherical_toroidal} is simply Green's function expansion \eqref{GF} for $\b r_1=0$. And for $n=1$, $m=0,1$, \eqref{reg_spherical_toroidal} and \eqref{irr_spherical_toroidal} are expansions of uniform fields and point dipoles at the origin. Some low orders of the coefficients are
\begin{align}
c_{k0}^0&=2, \qquad 
&s_{k1}^0&=8k,  \nonumber\\
c_{k2}^0&=8k^2+2, \qquad
&s_{k3}^0&=\frac{16}{9}(4k^3+5k), \nonumber\\
c_{k1}^1&=-4k^2+1, \qquad
&c_{k1}^{-1}&=-4. 
\end{align}

\subsection{Explicit forms of spherical-toroidal expansion coefficients} \label{sec coefs}
While the recurrence \eqref{c s rec} is sufficient to accurately calculate the coefficients $c_{nk}^m$ and $s_{nk}^m$, we here investigate more equivalent explicit forms and properties.
First, Erofeenko \cite{erofeenko1983addition} deduced that these were related to the coefficients for expressing a product of trig functions as a series of powers of the tangent. Essentially they had
\begin{align}
 2^{m+1}\cos^{m+1}\!\theta\sin^m\!\theta \cos 2n\theta =& \sum_{k=m:2}^\infty c_{nk}^{-m} P_k^m(0) \tan^k\theta \label{cnkm tan} \\
-2^{m+1}\cos^{m+1}\!\theta\sin^m\!\theta \sin 2n\theta =& \!\!\sum_{k=m+1:2}^\infty\!\!\! s_{nk}^{-m}P_{k+1}^{m}(0)\tan^k\theta \label{snkm tan}
\end{align}
where the notation $\sum_{k=m:2}^\infty$ means that $k$ rises in steps of 2.
From this they derived explicit forms of the coefficients, and defined $h_{nk}^m$ which dealt with $\Psi_n^{mc}$ and $\Psi_n^{ms}$ simultaneously:
\begin{align}
h_{nk}^m=&(-)^k[c_{nk}^mP_k^{-m}(0) + i s_{nk}^mP_{k+1}^{-m}(0)] \nonumber\\
	    =&(-)^k2\frac{\Gamma(n+m+1/2)(k-m)!}{\Gamma(n-m+1/2)(k+m)!}\nonumber\\
	    &\times\bigg[\frac{2^{2m}m!}{(2m)!} + \sum_{p=1}^n \frac{(-)^p 2^{3p+2m}(p+m)!(n+p-1)!}{(2p)!(2p+2m)!(n-p)!}\nonumber\\
	    &\times\left(nP_{k-p}^{m+p}(0) - i(k+m)pP_{k-p}^{m+p-1}(0) \right)\bigg]. \label{hnkm 1983}
\end{align}
For $p>k$, the identity for Legendre functions $P_{-k-1}^m=P_k^m$ is implied. Also, in \cite{erofeenko1983addition} a dimensional factor of $a^{k+1}$ was included which we ignore here. 
$h_{nk}^m$ was also found to have a triangular recurrence over $n$ and $k$:
\begin{align}
2i(k-m)h_{n,k-1}^m &=\! \left(n-m+\frac{1}{2}\right)h_{n+1,k}^m - 2nh_{nk}^m \nonumber\\&+\left(n+m-\frac{1}{2}\right)h_{n-1,k}^m,
\end{align}
which was obtained by application of $\pd_z$ instead of $r\pd_r$ as in the derivation in section \ref{toroidal vs spherical appendix}. $h_{nk}^m$ was actually generalised to consider the centre of the spherical coordinates being translated up the $z$-axis - this is obtained essentially by replacing the 0 in $P_k^m(0)$ with some angle. These generalised coefficients could be used to obtain the $T$-matrix for the offset torus, allowing an analytic study of stacked tori, as considered in \cite{salhi2015toroidal}.\\

Also, by direct evaluation of the toroidal harmonic expansion \eqref{cos_toroidal_spherical} for $r<a$ on the $z$-axis, using $v=z/a$, we find:
\begin{align}
\frac{2^{m+1}}{(1+v^2)^{m+1/2}}T_n\bigg(\frac{1-v^2}{1+v^2}\bigg)=\sum_{k=m:2}^\infty c_{nk}^{-m} P_k^m(0) v^{k-m} \label{cnkm axis}
\end{align} 
where $T_n(\cos \theta)=\cos(n\theta)$ are the Chebyshev polynomials. \eqref{cnkm axis} can be seen to be equivalent to \eqref{cnkm tan} by substituting $v=\tan\theta$ and recognising that $1/\sqrt{1+v^2}=\cos\theta$ and $T_{2n}(x)=T_n(2x^2-1)$.  \\

We can also find the generating function. Adding \eqref{cnkm tan} and \eqref{snkm tan} gives
\begin{align}
2\sin^m 2\theta\cos\theta e^{i2n\theta} =& \sum_{k=m}^\infty h_{nk}^{-m} \tan^k\theta. \label{hnkm tan}
\end{align}
Substituting $v=\tan\theta$ gives the generating function for $h_{nk}^{-m}$:
\begin{align}
2^{m+1}\frac{(1-iv)^{n-m-1/2}}{(1+iv)^{n+m+1/2}}= \sum_{k=m}^\infty h_{nk}^{-m} v^{k-m} \label{hnkm GF}
\end{align}
By expanding this via the binomial series for the numerator and denominator separately, and multiplying these, it is possible to find $h_{nk}$ as a finite sum of binomial coefficients, then using an identity for the product of two binomial coefficients, we find
\begin{align}
h_{nk}^{-m}&= 2^{m+1}i^k\sum_{p=0}^{k-m}\binom{n-1/2+k-p}{k-p-m}\binom{n-m-1/2}{p}. \label{hnkm sum}
\end{align}
which is equivalent to \eqref{hnkm 1983}. 

By application of eq. 7.12, vol. 4 of \cite{gould2010} to \eqref{hnkm sum}, we also obtain

\begin{align}
h_{nk}^{-m}=\frac{2^{m+1}}{i^{m-k}}\sum_{p=0}^{\lfloor (k-m)/2\rfloor}\binom{2n}{k-m-2p}\binom{n+m+p-1/2}{p} \label{hnkm symmetric sum}
\end{align}

Using the generating function \eqref{hnkm GF}, it is possible to find two more recurrence relations:
\begin{align}
h_{n+1k}^{m} - i h_{n+1k-1}^{m} = h_{nk}^{m} + i h_{nk-1}^{m} \label{h rec 4term}
\end{align}
and a recurrence over $k$ only:
\begin{align}
(k+m+1)h_{nk+1}^m=2in h_{nk}^m - (k-m)h_{nk-1}^m. \label{h rec k}
\end{align}
\eqref{h rec k} is similar in form to the recurrence over $n$, \eqref{c s rec}, which for comparison we write here for $h_{nk}^m$:
\begin{align*}
\!\!\bigg(\!n-m+\frac{1}{2}\!\bigg) h_{n+1,k}^m \!= (2k+1)h_{nk}^m + \bigg(\!n+m-\frac{1}{2}\!\bigg)h_{n-1,k}^m 
\end{align*}
In fact if we define rational numbers $e_{nk}^m=i^{-k}h_{n+1/2,k}^m$ which may be computed using \eqref{hnkm sum}, then $e_{nk}^m$ is symmetric: $e_{nk}^m=e_{kn}^{-m}$. This symmetry is related to the fact that toroidal harmonics $\Delta P_{n-1/2}^m(\beta)e^{in\eta}$ and spherical harmonics $r^{-1/2}r^{\pm(n+1/2)}P_n^m(u)$ are of similar form, related by a symmetric coordinate transformation and index shift $n\leftrightarrow n+1/2$, as we will now show.
Defining $\alpha=e^{i\eta}$, $s=r/a$, we have the following algebraic relationships between toroidal and spherical coordinates:
\begin{align*}
\beta =&\frac{s+s^{-1}}{\sqrt{(s-s^{-1})^2+4u^2}} &
\alpha=&\sqrt{\frac{s-s^{-1}+2iu}{s-s^{-1}-2iu}}\\
u     =&\frac{\alpha-\alpha^{-1}}{\sqrt{(\alpha+\alpha^{-1})^2-4\beta^2}} &
s     =&\sqrt{\frac{\alpha+\alpha^{-1}+2\beta}{-\alpha-\alpha^{-1}+2\beta}}
\end{align*}
The transformation between the coordinate systems is symmetric and can be expressed by the substitutions $s\leftrightarrow i\alpha$, $u\leftrightarrow \beta$.
Also the separation constant $\Delta$ can be modified by a factor so that its reciprocal is equal to itself under coordinate transformation:
\begin{align*}
\sqrt{\frac{r}{2a}}\Delta =& \frac{[-(\alpha+\alpha^{-1})^2+4\beta^2]^{1/4}}{2} \equiv \bar\Delta\\
\bar\Delta^{-1} =& \frac{[(s-s^{-1})^2+4u^2]^{1/4}}{2}=\bar{\Delta}(i\alpha\rightarrow s, \beta\rightarrow u).
\end{align*}
with these coordinates, we can rewrite the toroidal-spherical expansions in a very symmetrical form (for example just looking at expansions involving regular spherical harmonics $\hat{S}_n^m$, (\ref{cos_toroidal_spherical}-\ref{reg_spherical_toroidal})): \\
\begin{align}
\bar\Delta \alpha^n P_{n-1/2}^m(\beta) =& \sum_{k=0}^\infty (-)^k h_{nk}^m s^{k+1/2} P_k^m(u) \label{symmetricexpn1}\\
\bar\Delta^{-1} s^{n+1/2} P_n^m(u) =& \sum_{k=0}^\infty \frac{\varepsilon_k}{2\pi} h_{kn}^{-m} \alpha^k Q_{k-1/2}^m(\beta) \label{symmetricexpn2}
\end{align}
these are closely related by $n\rightarrow n+1/2$, $k\leftrightarrow k-1/2$, $s\leftrightarrow i\alpha$, $u\leftrightarrow \beta$. The main difference is that \eqref{symmetricexpn1} is a series of $P_k^m$ while \eqref{symmetricexpn2} involves $Q_{k-1/2}^m$, which is required for the series to converge since $|u|\leq1$ and $\beta>1$.

Matlab codes are attached for computation of the Legendre functions, and the $T$-matrix  on both toroidal and spherical bases.

\bibliography{../libraryH}

\begin{thebibliography}{36}%
\makeatletter
\providecommand \@ifxundefined [1]{%
 \@ifx{#1\undefined}
}%
\providecommand \@ifnum [1]{%
 \ifnum #1\expandafter \@firstoftwo
 \else \expandafter \@secondoftwo
 \fi
}%
\providecommand \@ifx [1]{%
 \ifx #1\expandafter \@firstoftwo
 \else \expandafter \@secondoftwo
 \fi
}%
\providecommand \natexlab [1]{#1}%
\providecommand \enquote  [1]{``#1''}%
\providecommand \bibnamefont  [1]{#1}%
\providecommand \bibfnamefont [1]{#1}%
\providecommand \citenamefont [1]{#1}%
\providecommand \href@noop [0]{\@secondoftwo}%
\providecommand \href [0]{\begingroup \@sanitize@url \@href}%
\providecommand \@href[1]{\@@startlink{#1}\@@href}%
\providecommand \@@href[1]{\endgroup#1\@@endlink}%
\providecommand \@sanitize@url [0]{\catcode `\\12\catcode `\$12\catcode
  `\&12\catcode `\#12\catcode `\^12\catcode `\_12\catcode `\%12\relax}%
\providecommand \@@startlink[1]{}%
\providecommand \@@endlink[0]{}%
\providecommand \url  [0]{\begingroup\@sanitize@url \@url }%
\providecommand \@url [1]{\endgroup\@href {#1}{\urlprefix }}%
\providecommand \urlprefix  [0]{URL }%
\providecommand \Eprint [0]{\href }%
\providecommand \doibase [0]{http://dx.doi.org/}%
\providecommand \selectlanguage [0]{\@gobble}%
\providecommand \bibinfo  [0]{\@secondoftwo}%
\providecommand \bibfield  [0]{\@secondoftwo}%
\providecommand \translation [1]{[#1]}%
\providecommand \BibitemOpen [0]{}%
\providecommand \bibitemStop [0]{}%
\providecommand \bibitemNoStop [0]{.\EOS\space}%
\providecommand \EOS [0]{\spacefactor3000\relax}%
\providecommand \BibitemShut  [1]{\csname bibitem#1\endcsname}%
\let\auto@bib@innerbib\@empty
\bibitem [{\citenamefont {Waterman}(1965)}]{waterman1965matrix}%
  \BibitemOpen
  \bibfield  {author} {\bibinfo {author} {\bibfnamefont {P.}~\bibnamefont
  {Waterman}},\ }\href@noop {} {\bibfield  {journal} {\bibinfo  {journal}
  {Proceedings of the IEEE}\ }\textbf {\bibinfo {volume} {53}},\ \bibinfo
  {pages} {805} (\bibinfo {year} {1965})}\BibitemShut {NoStop}%
\bibitem [{\citenamefont {Mishchenko}\ \emph {et~al.}(2002)\citenamefont
  {Mishchenko}, \citenamefont {Travis},\ and\ \citenamefont
  {Lacis}}]{Mishchenko2002}%
  \BibitemOpen
  \bibfield  {author} {\bibinfo {author} {\bibfnamefont {M.~I.}\ \bibnamefont
  {Mishchenko}}, \bibinfo {author} {\bibfnamefont {L.~D.}\ \bibnamefont
  {Travis}}, \ and\ \bibinfo {author} {\bibfnamefont {A.~A.}\ \bibnamefont
  {Lacis}},\ }\href@noop {} {\emph {\bibinfo {title} {Scattering, absorption,
  and emission of light by small particles}}}\ (\bibinfo  {publisher}
  {Cambridge university press},\ \bibinfo {year} {2002})\BibitemShut {NoStop}%
\bibitem [{\citenamefont {Farafonov}\ \emph {et~al.}(2016)\citenamefont
  {Farafonov}, \citenamefont {Il'in}, \citenamefont {Ustimov}, \citenamefont
  {Prokopjeva} \emph {et~al.}}]{farafonov2016analysis}%
  \BibitemOpen
  \bibfield  {author} {\bibinfo {author} {\bibfnamefont {V.}~\bibnamefont
  {Farafonov}}, \bibinfo {author} {\bibfnamefont {V.}~\bibnamefont {Il'in}},
  \bibinfo {author} {\bibfnamefont {V.}~\bibnamefont {Ustimov}}, \bibinfo
  {author} {\bibfnamefont {M.}~\bibnamefont {Prokopjeva}},  \emph {et~al.},\
  }\href@noop {} {\bibfield  {journal} {\bibinfo  {journal} {Journal of
  Quantitative Spectroscopy and Radiative Transfer}\ }\textbf {\bibinfo
  {volume} {178}},\ \bibinfo {pages} {176} (\bibinfo {year}
  {2016})}\BibitemShut {NoStop}%
\bibitem [{\citenamefont {Kyurkchan}\ \emph {et~al.}(1996)\citenamefont
  {Kyurkchan}, \citenamefont {Sternin},\ and\ \citenamefont
  {Shatalov}}]{kyurkchan1996singularities}%
  \BibitemOpen
  \bibfield  {author} {\bibinfo {author} {\bibfnamefont {A.~G.}\ \bibnamefont
  {Kyurkchan}}, \bibinfo {author} {\bibfnamefont {B.~Y.}\ \bibnamefont
  {Sternin}}, \ and\ \bibinfo {author} {\bibfnamefont {V.~E.}\ \bibnamefont
  {Shatalov}},\ }\href@noop {} {\bibfield  {journal} {\bibinfo  {journal}
  {Physics-Uspekhi}\ }\textbf {\bibinfo {volume} {39}},\ \bibinfo {pages}
  {1221} (\bibinfo {year} {1996})}\BibitemShut {NoStop}%
\bibitem [{\citenamefont {Love}(1972)}]{love1972dielectric}%
  \BibitemOpen
  \bibfield  {author} {\bibinfo {author} {\bibfnamefont {J.}~\bibnamefont
  {Love}},\ }\href@noop {} {\bibfield  {journal} {\bibinfo  {journal} {Journal
  of Mathematical Physics}\ }\textbf {\bibinfo {volume} {13}},\ \bibinfo
  {pages} {1297} (\bibinfo {year} {1972})}\BibitemShut {NoStop}%
\bibitem [{\citenamefont {Venkov}(2007)}]{venkov2007low}%
  \BibitemOpen
  \bibfield  {author} {\bibinfo {author} {\bibfnamefont {G.}~\bibnamefont
  {Venkov}},\ }\href@noop {} {\bibfield  {journal} {\bibinfo  {journal}
  {Journal of Computational Acoustics}\ }\textbf {\bibinfo {volume} {15}},\
  \bibinfo {pages} {181} (\bibinfo {year} {2007})}\BibitemShut {NoStop}%
\bibitem [{\citenamefont {Dutta}\ \emph {et~al.}(2008)\citenamefont {Dutta},
  \citenamefont {Ali}, \citenamefont {Brandl}, \citenamefont {Park},\ and\
  \citenamefont {Nordlander}}]{dutta2008plasmonic}%
  \BibitemOpen
  \bibfield  {author} {\bibinfo {author} {\bibfnamefont {C.~M.}\ \bibnamefont
  {Dutta}}, \bibinfo {author} {\bibfnamefont {T.~A.}\ \bibnamefont {Ali}},
  \bibinfo {author} {\bibfnamefont {D.~W.}\ \bibnamefont {Brandl}}, \bibinfo
  {author} {\bibfnamefont {T.-H.}\ \bibnamefont {Park}}, \ and\ \bibinfo
  {author} {\bibfnamefont {P.}~\bibnamefont {Nordlander}},\ }\href@noop {}
  {\bibfield  {journal} {\bibinfo  {journal} {The Journal of chemical physics}\
  }\textbf {\bibinfo {volume} {129}},\ \bibinfo {pages} {084706} (\bibinfo
  {year} {2008})}\BibitemShut {NoStop}%
\bibitem [{\citenamefont {Salhi}\ \emph {et~al.}(2015)\citenamefont {Salhi},
  \citenamefont {Passian},\ and\ \citenamefont {Siopsis}}]{salhi2015toroidal}%
  \BibitemOpen
  \bibfield  {author} {\bibinfo {author} {\bibfnamefont {M.}~\bibnamefont
  {Salhi}}, \bibinfo {author} {\bibfnamefont {A.}~\bibnamefont {Passian}}, \
  and\ \bibinfo {author} {\bibfnamefont {G.}~\bibnamefont {Siopsis}},\
  }\href@noop {} {\bibfield  {journal} {\bibinfo  {journal} {Physical Review
  A}\ }\textbf {\bibinfo {volume} {92}},\ \bibinfo {pages} {033416} (\bibinfo
  {year} {2015})}\BibitemShut {NoStop}%
\bibitem [{\citenamefont {Garapati}\ \emph {et~al.}(2017)\citenamefont
  {Garapati}, \citenamefont {Salhi}, \citenamefont {Kouchekian}, \citenamefont
  {Siopsis},\ and\ \citenamefont {Passian}}]{garapati2017poloidal}%
  \BibitemOpen
  \bibfield  {author} {\bibinfo {author} {\bibfnamefont {K.~V.}\ \bibnamefont
  {Garapati}}, \bibinfo {author} {\bibfnamefont {M.}~\bibnamefont {Salhi}},
  \bibinfo {author} {\bibfnamefont {S.}~\bibnamefont {Kouchekian}}, \bibinfo
  {author} {\bibfnamefont {G.}~\bibnamefont {Siopsis}}, \ and\ \bibinfo
  {author} {\bibfnamefont {A.}~\bibnamefont {Passian}},\ }\href@noop {}
  {\bibfield  {journal} {\bibinfo  {journal} {Physical Review B}\ }\textbf
  {\bibinfo {volume} {95}},\ \bibinfo {pages} {165422} (\bibinfo {year}
  {2017})}\BibitemShut {NoStop}%
\bibitem [{\citenamefont {Vafeas}(2016)}]{vafeas2016torusdipole}%
  \BibitemOpen
  \bibfield  {author} {\bibinfo {author} {\bibfnamefont {P.}~\bibnamefont
  {Vafeas}},\ }\href@noop {} {\bibfield  {journal} {\bibinfo  {journal}
  {Mathematical Methods in the Applied Sciences}\ }\textbf {\bibinfo {volume}
  {39}},\ \bibinfo {pages} {4268} (\bibinfo {year} {2016})}\BibitemShut
  {NoStop}%
\bibitem [{\citenamefont {Vafeas}\ \emph {et~al.}(2009)\citenamefont {Vafeas},
  \citenamefont {Perrusson},\ and\ \citenamefont
  {Lesselier}}]{vafeas2009spheroid}%
  \BibitemOpen
  \bibfield  {author} {\bibinfo {author} {\bibfnamefont {P.}~\bibnamefont
  {Vafeas}}, \bibinfo {author} {\bibfnamefont {G.}~\bibnamefont {Perrusson}}, \
  and\ \bibinfo {author} {\bibfnamefont {D.}~\bibnamefont {Lesselier}},\
  }\href@noop {} {\bibfield  {journal} {\bibinfo  {journal} {International
  Journal of Engineering Science}\ }\textbf {\bibinfo {volume} {47}},\ \bibinfo
  {pages} {372} (\bibinfo {year} {2009})}\BibitemShut {NoStop}%
\bibitem [{\citenamefont {Vafeas}\ \emph {et~al.}(2012)\citenamefont {Vafeas},
  \citenamefont {Papadopoulos},\ and\ \citenamefont
  {Lesselier}}]{vafeas2012twospheres}%
  \BibitemOpen
  \bibfield  {author} {\bibinfo {author} {\bibfnamefont {P.}~\bibnamefont
  {Vafeas}}, \bibinfo {author} {\bibfnamefont {P.~K.}\ \bibnamefont
  {Papadopoulos}}, \ and\ \bibinfo {author} {\bibfnamefont {D.}~\bibnamefont
  {Lesselier}},\ }\href@noop {} {\bibfield  {journal} {\bibinfo  {journal}
  {Journal of Applied Mathematics}\ }\textbf {\bibinfo {volume} {2012}}
  (\bibinfo {year} {2012})}\BibitemShut {NoStop}%
\bibitem [{\citenamefont {Maji{\'c}}\ \emph {et~al.}(2017)\citenamefont
  {Maji{\'c}}, \citenamefont {Gray}, \citenamefont {Augui{\'e}},\ and\
  \citenamefont {Le~Ru}}]{TmatESA2017}%
  \BibitemOpen
  \bibfield  {author} {\bibinfo {author} {\bibfnamefont {M.~R.}\ \bibnamefont
  {Maji{\'c}}}, \bibinfo {author} {\bibfnamefont {F.}~\bibnamefont {Gray}},
  \bibinfo {author} {\bibfnamefont {B.}~\bibnamefont {Augui{\'e}}}, \ and\
  \bibinfo {author} {\bibfnamefont {E.~C.}\ \bibnamefont {Le~Ru}},\ }\href@noop
  {} {\bibfield  {journal} {\bibinfo  {journal} {Journal of Quantitative
  Spectroscopy and Radiative Transfer}\ }\textbf {\bibinfo {volume} {200}},\
  \bibinfo {pages} {50} (\bibinfo {year} {2017})}\BibitemShut {NoStop}%
\bibitem [{\citenamefont {Maji{\'c}}\ and\ \citenamefont
  {Le~Ru}(2019)}]{majic2019quasistatic}%
  \BibitemOpen
  \bibfield  {author} {\bibinfo {author} {\bibfnamefont {M.~R.}\ \bibnamefont
  {Maji{\'c}}}\ and\ \bibinfo {author} {\bibfnamefont {E.~C.}\ \bibnamefont
  {Le~Ru}},\ }\href@noop {} {\bibfield  {journal} {\bibinfo  {journal} {Journal
  of Quantitative Spectroscopy and Radiative Transfer}\ }\textbf {\bibinfo
  {volume} {225}},\ \bibinfo {pages} {16} (\bibinfo {year} {2019})}\BibitemShut
  {NoStop}%
\bibitem [{\citenamefont {Majic}\ \emph {et~al.}(2019)\citenamefont {Majic},
  \citenamefont {Pratley}, \citenamefont {Schebarchov}, \citenamefont
  {Somerville}, \citenamefont {Augui{\'e}},\ and\ \citenamefont
  {Le~Ru}}]{majic2019approximate}%
  \BibitemOpen
  \bibfield  {author} {\bibinfo {author} {\bibfnamefont {M.}~\bibnamefont
  {Majic}}, \bibinfo {author} {\bibfnamefont {L.}~\bibnamefont {Pratley}},
  \bibinfo {author} {\bibfnamefont {D.}~\bibnamefont {Schebarchov}}, \bibinfo
  {author} {\bibfnamefont {W.~R.}\ \bibnamefont {Somerville}}, \bibinfo
  {author} {\bibfnamefont {B.}~\bibnamefont {Augui{\'e}}}, \ and\ \bibinfo
  {author} {\bibfnamefont {E.~C.}\ \bibnamefont {Le~Ru}},\ }\href@noop {}
  {\bibfield  {journal} {\bibinfo  {journal} {Physical Review A}\ }\textbf
  {\bibinfo {volume} {99}},\ \bibinfo {pages} {013853} (\bibinfo {year}
  {2019})}\BibitemShut {NoStop}%
\bibitem [{\citenamefont {Erofeenko}(1983)}]{erofeenko1983addition}%
  \BibitemOpen
  \bibfield  {author} {\bibinfo {author} {\bibfnamefont {V.~T.}\ \bibnamefont
  {Erofeenko}},\ }\href@noop {} {\bibfield  {journal} {\bibinfo  {journal}
  {Differentsial'nye Uravneniya}\ }\textbf {\bibinfo {volume} {19}},\ \bibinfo
  {pages} {1416} (\bibinfo {year} {1983})}\BibitemShut {NoStop}%
\bibitem [{\citenamefont {Shushkevich}(1998)}]{shushkevich1998electrostatic}%
  \BibitemOpen
  \bibfield  {author} {\bibinfo {author} {\bibfnamefont {G.~C.}\ \bibnamefont
  {Shushkevich}},\ }\href@noop {} {\bibfield  {journal} {\bibinfo  {journal}
  {Technical Physics}\ }\textbf {\bibinfo {volume} {43}},\ \bibinfo {pages}
  {743} (\bibinfo {year} {1998})}\BibitemShut {NoStop}%
\bibitem [{\citenamefont {Andrews}(2006)}]{Andrews2006}%
  \BibitemOpen
  \bibfield  {author} {\bibinfo {author} {\bibfnamefont {M.}~\bibnamefont
  {Andrews}},\ }\href {\doibase 10.1016/j.elstat.2005.11.005} {\bibfield
  {journal} {\bibinfo  {journal} {Journal of Electrostatics}\ }\textbf
  {\bibinfo {volume} {64}},\ \bibinfo {pages} {664} (\bibinfo {year}
  {2006})}\BibitemShut {NoStop}%
\bibitem [{\citenamefont {Farafonov}(2014)}]{farafonov2014rayleigh}%
  \BibitemOpen
  \bibfield  {author} {\bibinfo {author} {\bibfnamefont {V.}~\bibnamefont
  {Farafonov}},\ }\href@noop {} {\bibfield  {journal} {\bibinfo  {journal}
  {Optics and Spectroscopy}\ }\textbf {\bibinfo {volume} {117}},\ \bibinfo
  {pages} {923} (\bibinfo {year} {2014})}\BibitemShut {NoStop}%
\bibitem [{\citenamefont {Scharstein}\ and\ \citenamefont
  {Wilson}(2005)}]{Scharstein2005}%
  \BibitemOpen
  \bibfield  {author} {\bibinfo {author} {\bibfnamefont {R.~W.}\ \bibnamefont
  {Scharstein}}\ and\ \bibinfo {author} {\bibfnamefont {H.~B.}\ \bibnamefont
  {Wilson}},\ }\href@noop {} {\bibfield  {journal} {\bibinfo  {journal}
  {Electromagnetics}\ }\textbf {\bibinfo {volume} {25}},\ \bibinfo {pages} {1}
  (\bibinfo {year} {2005})}\BibitemShut {NoStop}%
\bibitem [{\citenamefont {Farafonov}\ and\ \citenamefont
  {Ustimov}(2015)}]{farafonov2015EBCM}%
  \BibitemOpen
  \bibfield  {author} {\bibinfo {author} {\bibfnamefont {V.~G.}\ \bibnamefont
  {Farafonov}}\ and\ \bibinfo {author} {\bibfnamefont {V.~I.}\ \bibnamefont
  {Ustimov}},\ }\href {\doibase 10.1134/S0030400X15120103} {\bibfield
  {journal} {\bibinfo  {journal} {Optics and Spectroscopy}\ }\textbf {\bibinfo
  {volume} {119}},\ \bibinfo {pages} {1022} (\bibinfo {year}
  {2015})}\BibitemShut {NoStop}%
\bibitem [{\citenamefont {Kuyucak}\ \emph {et~al.}(1998)\citenamefont
  {Kuyucak}, \citenamefont {Hoyles},\ and\ \citenamefont
  {Chung}}]{kuyucak1998analytical}%
  \BibitemOpen
  \bibfield  {author} {\bibinfo {author} {\bibfnamefont {S.}~\bibnamefont
  {Kuyucak}}, \bibinfo {author} {\bibfnamefont {M.}~\bibnamefont {Hoyles}}, \
  and\ \bibinfo {author} {\bibfnamefont {S.-H.}\ \bibnamefont {Chung}},\
  }\href@noop {} {\bibfield  {journal} {\bibinfo  {journal} {Biophysical
  journal}\ }\textbf {\bibinfo {volume} {74}},\ \bibinfo {pages} {22} (\bibinfo
  {year} {1998})}\BibitemShut {NoStop}%
\bibitem [{{\relax DLMF}()}]{NIST:DLMF}%
  \BibitemOpen
  {\relax DLMF},\ \href {http://dlmf.nist.gov/} {\enquote {\bibinfo {title}
  {{\it NIST Digital Library of Mathematical Functions}},}\ }\bibinfo
  {howpublished} {http://dlmf.nist.gov/, Release 1.0.16 of 2017-09-18},\
  \bibinfo {note} {f.~W.~J. Olver, A.~B. {Olde Daalhuis}, D.~W. Lozier, B.~I.
  Schneider, R.~F. Boisvert, C.~W. Clark, B.~R. Miller and B.~V. Saunders,
  eds.}\BibitemShut {Stop}%
\bibitem [{\citenamefont {Belevitch}\ and\ \citenamefont
  {Boersma}(1983)}]{belevitch1983torus}%
  \BibitemOpen
  \bibfield  {author} {\bibinfo {author} {\bibfnamefont {V.}~\bibnamefont
  {Belevitch}}\ and\ \bibinfo {author} {\bibfnamefont {J.}~\bibnamefont
  {Boersma}},\ }\href@noop {} {\bibfield  {journal} {\bibinfo  {journal}
  {Philips Journal of Research}\ }\textbf {\bibinfo {volume} {38}},\ \bibinfo
  {pages} {79} (\bibinfo {year} {1983})}\BibitemShut {NoStop}%
\bibitem [{\citenamefont {Avramov}\ \emph {et~al.}(1993)\citenamefont
  {Avramov}, \citenamefont {Ivanov},\ and\ \citenamefont
  {Zhelyazkov}}]{avramov1993high}%
  \BibitemOpen
  \bibfield  {author} {\bibinfo {author} {\bibfnamefont {K.}~\bibnamefont
  {Avramov}}, \bibinfo {author} {\bibfnamefont {T.}~\bibnamefont {Ivanov}}, \
  and\ \bibinfo {author} {\bibfnamefont {I.}~\bibnamefont {Zhelyazkov}},\
  }\href@noop {} {\bibfield  {journal} {\bibinfo  {journal} {Plasma physics and
  controlled fusion}\ }\textbf {\bibinfo {volume} {35}},\ \bibinfo {pages}
  {1787} (\bibinfo {year} {1993})}\BibitemShut {NoStop}%
\bibitem [{\citenamefont {Love}(1975)}]{love1975quasi}%
  \BibitemOpen
  \bibfield  {author} {\bibinfo {author} {\bibfnamefont {J.~D.}\ \bibnamefont
  {Love}},\ }\href@noop {} {\bibfield  {journal} {\bibinfo  {journal} {Journal
  of Plasma Physics}\ }\textbf {\bibinfo {volume} {14}},\ \bibinfo {pages} {25}
  (\bibinfo {year} {1975})}\BibitemShut {NoStop}%
\bibitem [{\citenamefont {Salhi}(2016)}]{salhi2016thesis}%
  \BibitemOpen
  \bibfield  {author} {\bibinfo {author} {\bibfnamefont {M.}~\bibnamefont
  {Salhi}},\ }\href@noop {} {\bibfield  {journal} {\bibinfo  {journal} {{PhD
  thesis, University of Tennessee}}\ } (\bibinfo {year} {2016})}\BibitemShut
  {NoStop}%
\bibitem [{\citenamefont {El-Shenawee}\ \emph {et~al.}(2009)\citenamefont
  {El-Shenawee}, \citenamefont {Macias}, \citenamefont {Baudrion},\ and\
  \citenamefont {Bachelot}}]{el2009torus}%
  \BibitemOpen
  \bibfield  {author} {\bibinfo {author} {\bibfnamefont {M.}~\bibnamefont
  {El-Shenawee}}, \bibinfo {author} {\bibfnamefont {D.}~\bibnamefont {Macias}},
  \bibinfo {author} {\bibfnamefont {A.~L.}\ \bibnamefont {Baudrion}}, \ and\
  \bibinfo {author} {\bibfnamefont {R.}~\bibnamefont {Bachelot}},\ }in\
  \href@noop {} {\emph {\bibinfo {booktitle} {Antennas and Propagation Society
  International Symposium, 2009. APSURSI'09. IEEE}}}\ (\bibinfo {organization}
  {IEEE},\ \bibinfo {year} {2009})\ pp.\ \bibinfo {pages} {1--4}\BibitemShut
  {NoStop}%
\bibitem [{\citenamefont {Mary}\ \emph {et~al.}(2007)\citenamefont {Mary},
  \citenamefont {Koller}, \citenamefont {Hohenau}, \citenamefont {Krenn},
  \citenamefont {Bouhelier},\ and\ \citenamefont {Dereux}}]{mary2007optical}%
  \BibitemOpen
  \bibfield  {author} {\bibinfo {author} {\bibfnamefont {A.}~\bibnamefont
  {Mary}}, \bibinfo {author} {\bibfnamefont {D.}~\bibnamefont {Koller}},
  \bibinfo {author} {\bibfnamefont {A.}~\bibnamefont {Hohenau}}, \bibinfo
  {author} {\bibfnamefont {J.}~\bibnamefont {Krenn}}, \bibinfo {author}
  {\bibfnamefont {A.}~\bibnamefont {Bouhelier}}, \ and\ \bibinfo {author}
  {\bibfnamefont {A.}~\bibnamefont {Dereux}},\ }\href@noop {} {\bibfield
  {journal} {\bibinfo  {journal} {Physical Review B}\ }\textbf {\bibinfo
  {volume} {76}},\ \bibinfo {pages} {245422} (\bibinfo {year}
  {2007})}\BibitemShut {NoStop}%
\bibitem [{\citenamefont {Le~Ru}\ and\ \citenamefont
  {Etchegoin}(2009)}]{LeRu2009}%
  \BibitemOpen
  \bibfield  {author} {\bibinfo {author} {\bibfnamefont {E.}~\bibnamefont
  {Le~Ru}}\ and\ \bibinfo {author} {\bibfnamefont {P.}~\bibnamefont
  {Etchegoin}},\ }\href@noop {} {\emph {\bibinfo {title} {Principles of
  Surface-Enhanced Raman Spectroscopy: and related plasmonic effects}}}\
  (\bibinfo  {publisher} {Elsevier},\ \bibinfo {year} {2009})\BibitemShut
  {NoStop}%
\bibitem [{\citenamefont {Weston}(1960)}]{weston1960toroidal}%
  \BibitemOpen
  \bibfield  {author} {\bibinfo {author} {\bibfnamefont {V.}~\bibnamefont
  {Weston}},\ }\href@noop {} {\bibfield  {journal} {\bibinfo  {journal}
  {Journal of Mathematics and Physics}\ } (\bibinfo {year} {1960})}\BibitemShut
  {NoStop}%
\bibitem [{\citenamefont {Mason}\ and\ \citenamefont
  {Handscomb}(2002)}]{mason2002chebyshev}%
  \BibitemOpen
  \bibfield  {author} {\bibinfo {author} {\bibfnamefont {J.~C.}\ \bibnamefont
  {Mason}}\ and\ \bibinfo {author} {\bibfnamefont {D.~C.}\ \bibnamefont
  {Handscomb}},\ }\href@noop {} {\emph {\bibinfo {title} {Chebyshev
  polynomials}}}\ (\bibinfo  {publisher} {Chapman and Hall/CRC},\ \bibinfo
  {year} {2002})\BibitemShut {NoStop}%
\bibitem [{\citenamefont {Morse}\ and\ \citenamefont
  {Feshbach}(1953)}]{Morse1953}%
  \BibitemOpen
  \bibfield  {author} {\bibinfo {author} {\bibfnamefont {P.~M.}\ \bibnamefont
  {Morse}}\ and\ \bibinfo {author} {\bibfnamefont {H.}~\bibnamefont
  {Feshbach}},\ }\href@noop {} {\enquote {\bibinfo {title} {{Methods of
  Theoretical Physics [Part 1 Chaps 1-8] 1953.pdf}},}\ } (\bibinfo {year}
  {1953})\BibitemShut {NoStop}%
\bibitem [{\citenamefont {Cohl}\ and\ \citenamefont
  {Tohline}(1999)}]{Cohl1999}%
  \BibitemOpen
  \bibfield  {author} {\bibinfo {author} {\bibfnamefont {H.~S.}\ \bibnamefont
  {Cohl}}\ and\ \bibinfo {author} {\bibfnamefont {J.~E.}\ \bibnamefont
  {Tohline}},\ }\href {\doibase 10.1086/308062} {\bibfield  {journal} {\bibinfo
   {journal} {The Astrophysical Journal}\ }\textbf {\bibinfo {volume} {527}},\
  \bibinfo {pages} {86} (\bibinfo {year} {1999})}\BibitemShut {NoStop}%
\bibitem [{\citenamefont {Hamed}(2014)}]{Hamed2014}%
  \BibitemOpen
  \bibfield  {author} {\bibinfo {author} {\bibfnamefont {S.~M.~A.}\
  \bibnamefont {Hamed}},\ }\href {\doibase 10.1109/TAP.2014.2329002} {\bibfield
   {journal} {\bibinfo  {journal} {IEEE Transactions on Antennas and
  Propagation}\ }\textbf {\bibinfo {volume} {62}},\ \bibinfo {pages} {4434}
  (\bibinfo {year} {2014})}\BibitemShut {NoStop}%
\bibitem [{\citenamefont {Gould}(1990)}]{gould2010}%
  \BibitemOpen
  \bibfield  {author} {\bibinfo {author} {\bibfnamefont {H.~W.}\ \bibnamefont
  {Gould}},\ }\href {https://www.math.wvu.edu/~gould/} {\bibfield  {journal}
  {\bibinfo  {journal} {unpuplished notes, edited by Jocelyn Quaintance}\
  }\textbf {\bibinfo {volume} {4}} (\bibinfo {year} {1945-1990})}\BibitemShut
  {NoStop}%
\end{thebibliography}%
\end{document}